\documentclass[acmsmall,screen,authorversion,nonacm,timestamp=false]{acmart}

\usepackage{xcolor, nicefrac}
\usepackage{tcolorbox}
\usepackage{multirow}
\usepackage{multicol}
\usepackage{subcaption}
\usepackage{wrapfig}
\usepackage{footnote}
\usepackage{tablefootnote}
\usepackage{siunitx}
\usepackage{hyperref}
\usepackage{breakurl}

\usepackage{tikz}
\usetikzlibrary{calc}
\usetikzlibrary{positioning}
\usetikzlibrary{patterns}
\usepackage{listings}
\usepackage{multicol}
\lstdefinestyle{mystyle}{
    numbers=left,
    basicstyle=\ttfamily\scriptsize,
    keywordstyle=\bfseries\color{blue},
    frame=single,
    xleftmargin=6mm,
    xrightmargin=2mm,
    commentstyle=\color{gray},
}
\lstdefinelanguage{Ibis}
{
    language=C++,
    morekeywords=
    {
        uint32,
        uint8,
        int32,
        thread_rate,
        float32,
        atomic,
        wait_for,
        schedule,
        decltype
    },
    classoffset=1,
    morekeywords={
        ordering_example,
        static_count_if,
        pipelined_for,
        lock,
        dynamic_count_if,
        replicated_count_if,
        map,
        reduce,
        map_reduce,
        spec_pipelined_for,
        pipelined_do,
        spec_count_if,
        body
    },
    keywordstyle=\bfseries,
    classoffset=0
}
\lstdefinelanguage{HLS}
{
    language=C++,
    otherkeywords=
    {
        hls::stream
    },
    classoffset=1,
    morekeywords={
        InsertionCell,
        InsertionSort
    },
    keywordstyle=\bfseries,
    classoffset=0
}

\makeatletter
\tikzset{
 queue/.style={outer sep=0mm,
 pattern=horizontal lines,
 minimum width=1cm,
 minimum height=1cm,
 append after command={
  \pgfextra{
  \draw(\tikzlastnode.north west) -- (\tikz@last@fig@name.south west);
  \draw(\tikzlastnode.north west) -- (\tikz@last@fig@name.north east);
  \draw(\tikzlastnode.north east) -- (\tikz@last@fig@name.south east);
  \draw(\tikzlastnode.south west) -- (\tikz@last@fig@name.south east);
  \node[align=center,left=0.0mm of \tikz@last@fig@name] {#1};}
  }
 }
}
\makeatother

\newcommand{\langname}{Kanagawa}
\newcommand{\execname}{Wavefront Threading}

\newcommand{\consistname}{Wavefront Consistency}
\newcommand{\eg}{{\em e.g.},}
\newcommand{\ie}{{\em i.e.},}

\AtBeginDocument{%
  }

\begin{document}

\def\HS{\hspace{\fontdimen2\font}}

\title{Wavefront Threading Enables Effective High-Level Synthesis}

\author{Blake Pelton}
\orcid{0009-0009-9204-1873}
\affiliation{%
  \institution{Microsoft}
  \city{Redmond}
  \country{USA}
}
\email{blakep@microsoft.com}

\author{Adam Sapek}
\orcid{0009-0006-8599-0171}
\affiliation{%
  \institution{Microsoft}
  \city{Redmond}
  \country{USA}
}
\email{adamsap@microsoft.com}

\author{Ken Eguro}
\orcid{0000-0001-5797-3661}
\affiliation{%
  \institution{Microsoft}
  \city{Redmond}
  \country{USA}
}
\email{eguro@microsoft.com}

\author{Daniel Lo}
\orcid{0009-0002-6504-9078}
\affiliation{%
  \institution{Microsoft}
  \city{Redmond}
  \country{USA}
}
\email{dlo@microsoft.com}

\author{Alessandro Forin}
\orcid{0000-0003-2902-2337}
\affiliation{%
  \institution{Microsoft}
  \city{Redmond}
  \country{USA}
}
\email{sandrof@microsoft.com}

\author{Matt Humphrey}
\orcid{0009-0006-5339-1297}
\affiliation{%
  \institution{Microsoft}
  \city{Redmond}
  \country{USA}
}
\email{mhumphr@microsoft.com}

\author{Jinwen Xi}
\orcid{0009-0003-6681-1674}
\affiliation{%
  \institution{Microsoft}
  \city{Redmond}
  \country{USA}
}
\email{jixi@microsoft.com}

\author{David Cox}
\orcid{0009-0008-1944-4337}
\affiliation{%
  \institution{Microsoft}
  \city{Redmond}
  \country{USA}
}
\email{coxdavid@microsoft.com}

\author{Rajas Karandikar}
\orcid{0009-0000-2122-6373}
\authornote{Author was at Purdue University during their contribution to this work. They are now at Microsoft.}
\affiliation{%
  \institution{Microsoft}
  \city{Redmond}
  \country{USA}
}
\email{rakarand@microsoft.com}

\author{Johannes de Fine Licht}
\orcid{0000-0002-1500-7411}
\authornote{Author was at ETH Zurich during their contribution to this work. They are now at NextSilicon.}
\affiliation{%
  \institution{ETH Zurich}
  \city{Zurich}
  \country{Switzerland}
}
\email{definelicht@inf.ethz.ch}

\author{Evgeny Babin}
\orcid{0009-0000-9750-6048}
\affiliation{%
  \institution{Microsoft}
  \city{Redmond}
  \country{USA}
}
\email{evgenybabin@microsoft.com}

\author{Adrian Caulfield}
\orcid{0000-0002-1707-4693}
\affiliation{%
  \institution{Microsoft}
  \city{Redmond}
  \country{USA}
}
\email{acaulfie@microsoft.com}

\author{Doug Burger}
\orcid{0009-0006-6588-6596}
\affiliation{%
  \institution{Microsoft}
  \city{Redmond}
  \country{USA}
}
\email{dburger@microsoft.com}
\renewcommand{\shortauthors}{Pelton, Sapek, Eguro, Lo, Forin, Humphrey, Xi, Cox, Karandikar, de Fine Licht, Babin, Caulfield, and Burger}

\begin{abstract}

Digital systems are growing in importance and computing hardware is growing more heterogeneous. 
Hardware design, however, remains laborious and expensive, in part due to the limitations of conventional hardware description languages (HDLs) like VHDL and Verilog. A longstanding research goal has been programming hardware like software, with high-level languages that can generate efficient hardware designs.  This paper describes \langname{}, a language that takes a new approach to combine the programmer productivity benefits of traditional High-Level Synthesis (HLS) approaches with the expressibility and hardware efficiency of Register-Transfer Level (RTL) design.  The language's concise syntax, matched with a hardware design-friendly execution model, permits a relatively simple toolchain to map high-level code into efficient hardware implementations.

\end{abstract}

\maketitle

\section{Introduction} \label{sec:intro}

The challenge of efficiently designing hardware using high-level programming languages has long been a focus of research. Hardware design, whether for flexible logic on Field-Programmable Gate Arrays (FPGAs) or hardened Application-Specific Integrated Circuits (ASICs), is dominated by Register-Transfer Level (RTL) Hardware Description Languages (HDLs) like Verilog and VHDL. These languages excel in expressiveness within the hardware domain, but suffer from a low level of abstraction that limits productivity. Designers must explicitly translate algorithmic intentions into low-level logic. While high-level HDLs such as Bluespec \cite{bluespec-2004} and Chisel \cite{chisel-dac12} provide richer abstractions, they remain largely tethered to the low-level hardware programming paradigm.

High-Level Synthesis (HLS) offers the promise of designing hardware using an imperative programming paradigm that has software development-like productivity. However, to date, HLS solutions have not seen widespread adoption due to challenging limitations. Automatically mapping sequential imperative code to efficient parallel hardware has proven to be a complex compiler design problem. Traditional HLS tools often suffer from unpredictable performance and lack language features to express various design choices. To circumvent these limitations, users often must resort to steering compiler decisions using hints and pragmas which negate the benefits of programming at a higher level of abstraction, limiting code reuse and composability.

\langname{} is a high-level, imperative programming language developed specifically for hardware design.  \langname{} relies on a new execution model for imperative code called \execname{}.  This execution model enables straightforward and predictable mapping to efficient parallel hardware. \langname{} offers expressiveness rivaling RTL, with the high level of abstraction and productivity of software development. Synthesized \langname{} designs compare in quality to those manually written by experienced designers in RTL. \langname{} programmers have leveraged its versatility to create a wide range of significant hardware designs, some of which are in large-scale production in Microsoft Azure.  \langname{} is sufficiently expressive to develop a full RISC-V core.

In the subsequent sections of this paper, we will illustrate how the \langname{} execution model addresses the challenges of previous HLS tools, describe the \langname{} language, explain how source code is mapped to hardware, and provide a quantitative assessment of \langname{} versus production-quality RTL designs.        
\section{Motivational Example}
\label{sec:motivation}
HLS tools commonly pipeline loops (\ie{} execute multiple iterations in parallel) to achieve high performance.  Figure \ref{CountIfC} shows the \texttt{CountIf Histogram} loop from \citet{dai-fpga17}, written in C.  A similar example appears in \citet{dynamic-hls-2018}.  When pipelining the loop there is a read-after-write hazard on \texttt{hist[m]}\footnote{Iteration \texttt{i} writes on line 6 and iteration \texttt{i+1} reads on line 5}.  How often this hazard must be addressed is data dependent.  Assume that the histogram read, floating-point addition, and histogram write have a combined latency of \(L\) clock cycles.  There are multiple options for handling this hazard, including:

\begin{enumerate}
\item Statically schedule the loop so that different iterations do not concurrently execute the read-modify-write. For example, configure the pipeline initiation interval \cite{Rau-1981} to \(L\). This approach provides limited throughput regardless of hazard frequency, but has the lowest resource usage. Figure \ref{CountIfExeStatic} illustrates a statically scheduled execution for initiation interval=2.
\item Dynamically detect hazards and stall when necessary. This option can improve best case throughput (\ie{} when hazards are rare) at the cost of extra logic resources.  Figure \ref{CountIfExeDynamic} illustrates a dynamically scheduled execution.
\item  Interleave the computation among \(L\) intermediate histograms. Add a post-processing step to combine the histograms and produce the final result\footnote{Assume the developer has determined floating-point addition in this application can be treated as associative.}. This option provides high throughput independent of hazard frequency but requires more memory and logic resources.
\item Speculatively read and modify \texttt{hist} before previous writes have completed.  Before writing to \texttt{hist}, check to see if there was a hazard and restart the iteration if necessary. This option also improves best case throughput and has resource costs comparable to option 2.
\end{enumerate}

\begin{figure*}[t]
  \begin{minipage}[t]{.4\linewidth}
  \begin{lstlisting}[language=C]
for (i=0; i<N; ++i) {
  int m = feature[i];
  float wt = weight[i];
  if (m>THRESHOLD) {
    float x = hist[m];
    hist[m] = x + wt;
  }
}
  \end{lstlisting}
  \caption{\texttt{CountIf Histogram} in C.}
  \label{CountIfC}
\end{minipage}
    \begin{tikzpicture}[
    every node/.style={draw},text depth=.25ex,text height=.25em,text width=2.25em, font=\scriptsize,
    i0node/.style={fill=red!20},
    i1node/.style={fill=blue!20},
    i2node/.style={fill=green!20},
    column 1/.style={nodes={text width=5em}},
    row 6/.style={nodes={font=\scriptsize \bfseries}},]
    \matrix [draw=none,row sep=0mm,column sep=0mm]
    {
    \node{\texttt{m=feature[i];}};    & \node[i0node] {I0\{\}};   & \node {};                     & \node[i1node] {I1\{\}};     & \node {};                  & \node [i2node] {I2\{\}};       & \node {};                   & \node {};                 & \node {};                         \\
    \node{\texttt{wt=weight[i];}};    & \node {};                 & \node[i0node] {I0\{m:5\}};    & \node{};                    & \node[i1node] {I1\{m:7\}}; & \node {};                      & \node[i2node] {I2\{m:7\}};  & \node {};                 & \node {};                         \\
    \node{\texttt{x=hist[m];}};       & \node {};                 & \node{};                      & \node[i0node] {I0\{m:5\}};  & \node {};                  & \node[i1node] {I1\{m:7\}};     & \node {};                   & \node[i2node] {I2\{m:7\}};& \node {};                         \\
    \node{\texttt{hist[m]=x+wt;}};    & \node {};                 & \node{};                      & \node{};                    & \node[i0node] {I0\{m:5\}}; & \node{};                       & \node[i1node] {I1\{m:7\}};  & \node {};                 & \node[i2node] {I2\{m:7\}};        \\
    \node{Time};                      & \node {0};                & \node {1};                    & \node {2};                  & \node {3};                 & \node {4};                     & \node {5};                  & \node {6};                & \node {7};                        \\
    };  
    \end{tikzpicture}
    \caption{An execution of \texttt{CountIf Histogram} using static scheduling with \(II = 2\).  Three loop iterations (\(I0\), \(I1\), and \(I2\)) are shown with distinct colors.  The value of \texttt{m} is shown at each iteration.}
    \label{CountIfExeStatic}
    \begin{tikzpicture}[
    every node/.style={draw},text depth=.25ex,text height=.25em,text width=2.25em,font=\scriptsize,
    i0node/.style={fill=red!20},
    i1node/.style={fill=blue!20},
    i2node/.style={fill=green!20},
    i2stal/.style={pattern=north west lines, pattern color=green!50},
    column 1/.style={nodes={text width=5em}},
    row 6/.style={nodes={font=\scriptsize \bfseries}},]
    \matrix [draw=none,row sep=0mm,column sep=0mm]
    {
    \node{\texttt{m=feature[i];}};    & \node[i0node] {I0\{\}};   & \node[i1node] {I1\{\}};       & \node[i2node] {I2\{\}};     & \node{};                   & \node {};                      & \node {};                     & \node {};                       & \node {};                     \\
    \node{\texttt{wt=weight[i];}};    & \node {};                 & \node[i0node] {I0\{m:5\}};    & \node[i1node] {I1\{m:7\}};  & \node[i2node] {I2\{m:7\}}; & \node{};                       & \node{};                      & \node {};                       & \node{};                      \\
    \node{\texttt{x=hist[m];}};       & \node {};                 & \node{};                      & \node[i0node] {I0\{m:5\}};  & \node[i1node] {I1\{m:7\}}; & \node[i2stal] {I2\{m:7\}};     & \node [i2node] {I2\{m:7\}};   & \node {};                       & \node{};                      \\
    \node{\texttt{hist[m]=x+wt;}};    & \node {};                 & \node{};                      & \node{};                    & \node[i0node] {I0\{m:5\}}; & \node[i1node] {I1\{m:7\}};     & \node {};                     & \node [i2node] {I2\{m:7\}};     & \node {};                     \\
    \node{Time};                      & \node {0};                & \node {1};                    & \node {2};                  & \node {3};                 & \node {4};                     & \node {5};                    & \node {6};                      & \node {7};                    \\
    };
    \end{tikzpicture}
    \caption{An execution of \texttt{CountIf Histogram} using dynamic scheduling.  The dashed cell at time 4 shows when \(I2\) must stall to resolve the read-after-write hazard.}%
    \label{CountIfExeDynamic}
\end{figure*}%

In imperative languages supported by traditional HLS tools (\eg{} C/C++), it is difficult to fully express all these design choices because the concurrency of loop iterations is not explicit and must be inferred by the compiler. For example, option 3 could be partially expressed by scaling the \texttt{hist} array size by \(L\) and adding a post-processing loop, but the programmer would have to rely on the compiler to recognize that it is safe to run up to \(L\) iterations of the primary loop concurrently or add tool-specific hints or pragmas.

Commercial HLS tools typically choose option 1, at the cost of throughput when hazards are rare.  Research into dynamic HLS \cite{dynamic-hls-2018} proposes ways to achieve option 2 when hazards are common, at the cost of hardware resources.  Other research \cite{spec-2020, spec-2022} proposes automated generation of speculative hardware (option 4, possibly with hints about the likely outcome of a conditional).  Hybrid approaches are also possible, such as combining speculation with dynamic HLS \cite{spec-19}.

The problem is fundamental to existing HLS approaches: sequential imperative code does not include enough context information for a tractable compiler to reliably generate efficient parallel hardware.  Prior work \cite{morvan-2011, splitting-2016, dai-fpga17, liu-cadics17, cheng-2021} has shown that HLS tools struggle to generate efficient hardware while maintaining sequential semantics for basic control flow compositions such as nested/sequential/irregular loops and for non-trivial loop-carried dependencies.  

\section{\langname{} Language}
\label{sec:language}

\langname{} has a threaded execution model and associated memory consistency model that facilitates translation to hardware.  Concurrency is exposed explicitly using threads.  The language and execution model support fine-grained threads running imperative code.  In this section, we describe the semantics of the execution and memory models using \texttt{CountIf Histogram} to illustrate them in practice.  Three of the \texttt{CountIf Histogram} implementation options are discussed in this section and the fourth is described in Section \ref{sec:complex}.

\subsection{\execname{}}
\langname{} threads are lightweight, which encourages the expression of fine-grained concurrency, similar to how low per-thread overhead in GPU programming \cite{cuda-2008} facilitates the development of highly concurrent programs.  It is common, for example, to execute each iteration of a loop using a different thread. The key property of the \execname{} model that reduces the complexity of reasoning about highly concurrent code is that threads maintain relative order when executing imperative code. If thread \(T0\) starts executing code \textit{ahead of} thread \(T1\), \(T0\) will remain ahead of \(T1\) throughout execution, unless the ordering is explicitly relaxed using dedicated control flow constructs (Section \ref{sec:ControlFlow}).

Preserving thread order during execution not only simplifies reasoning about correctness, but also eliminates the need for explicit synchronization in many cases. Consider the example of \texttt{CountIf Histogram}, which can be challenging to parallelize in a typical multi-threaded language if floating-point addition cannot be treated as associative. Preserving the order of additions would require a complex synchronization mechanism such as ticket locks \cite{ticket-1991}, and the time spent acquiring locks would likely dominate the runtime. With \execname{}, no synchronization is needed because thread ordering guarantees that the additions occur in the correct order. Another consequence of \execname{} is that, during the execution of a function by a group of threads, each thread's position within the group is well defined. This information can often be used in lieu of synchronization. For instance, the first thread in a group can be designated to reset shared state to an initial value, while the final thread can handle post-processing tasks.

The term "wavefront" reflects how a computation progresses in time, similar to Wavefront parallelism \cite{wavefront-1982, chamberlain1999array, skewing-1986}. As illustrated in Figure \ref{OrderedExecExample}, each thread can be conceptualized as a wave flowing through the computation from top-left to bottom-right.

\subsection{Shared State}
State in \langname{} is stored in named variables, with a strict separation between thread-local and shared state.  Threads communicate through shared variables (class members or static local variables) and use synchronization mechanisms from the language and libraries to ensure correctness.  There are no pointers, so aliasing is impossible. 

\subsection{\consistname{}}

Because \langname{} models synchronous digital circuits, memory\footnote{The term "memory" has different connotations in the programming languages and hardware design fields. In this work, we use the terms "memory" and "shared variable" synonymously to refer to state shared between threads.} consistency is defined in terms of a global clock, where each access to memory occurs at a particular clock cycle, and multiple accesses can occur at the same time (\ie{} the same clock cycle).

The model is similar to sequential consistency \cite{lamport-1997} in that any valid execution must behave as if all memory accesses performed by a thread are issued in the order defined by the program. \consistname{} further restricts the set of valid executions based on thread ordering. If thread \(T0\) is ahead of thread \(T1\) and both threads execute the same memory access operation, then the model defines that the thread \(T0\) will perform the memory access before thread \(T1\).

\begin{figure*}[t]
  \begin{minipage}[t]{.6\linewidth}
    \begin{lstlisting}[language=Ibis]
uint32 a = x; // Read from shared variable x
uint32 b = y; // Read from shared variable y
x = a+b;      // Write to shared variable x
y = a*2;      // Write to shared variable y
    \end{lstlisting}
  \end{minipage}
\begin{tikzpicture}[
every node/.style={draw},text depth=.25ex,text height=0.25em,text width=2.75em,font=\scriptsize,
t0node/.style={fill=red!20},
t0upda/.style={fill=red!20},
t1node/.style={fill=blue!20},
t1stal/.style={pattern=north west lines, pattern color=blue!50},
t1upda/.style={fill=blue!20},
t2node/.style={fill=green!20},
t2stal/.style={pattern=north west lines, pattern color=green!50},
t2upda/.style={fill=green!20},
column 1/.style={nodes={text width=3em}},
row 7/.style={nodes={font=\scriptsize \bfseries}},]
\matrix [draw=none,row sep=0mm,column sep=0mm]
{
\node{\texttt{a=x;}};     & \node[t0node] {T0\{\}}; & \node[t1node] {T1\{\}};     & \node[t2stal] {T2\{\}};           & \node[t2node] {T2\{\}};    & \node {};                        & \node {};                         & \node {};                     & \node {};  \\
\node{\texttt{b=y;}};     & \node {};               & \node[t0node] {T0\{a:2\}};  & \node[t1stal] {T1\{a:2\}};        & \node[t1node] {T1\{a:2\}}; & \node[t2node] {T2\{a:5\}};       & \node {};                         & \node {};                     & \node {}; \\
\node{\texttt{x=a+b;}};   & \node {};               & \node{};                    & \node[t0node] {T0\{a:2,b:3\}};    & \node {};                  & \node[t1node] {T1\{a:2,b:3\}};   & \node[t2node] {T2\{a:5,b:4\}};    & \node {};                     & \node {}; \\
\node{\texttt{y=a*2;}};   & \node {};               & \node{};                    & \node {};                         & \node[t0node] {T0\{a:2\}}; & \node {};                        & \node [t1node] {T1\{a:2\}};       & \node[t2node] {T2\{a:5\}};    & \node {};\\[2mm]
\node{\texttt{x}};          & \node {2};              & \node{2};                   & \node {2};                        & \node[t0upda] {5};    & \node {5};                       & \node[t1upda]  {5};          & \node[t2upda] {9};       & \node {9};\\
\node{\texttt{y}};          & \node {3};              & \node{3};                   & \node {3};                        & \node {3};                 & \node[t0upda] {4};          & \node  {4};                       & \node[t1upda] {4};       & \node[t2upda] {10};\\[2mm]
\node{Time};                & \node {0};              & \node{1};                   & \node {2};                        & \node {3};                 & \node {4};                       & \node  {5};                       & \node {6};                    & \node {7};\\
};
\end{tikzpicture}
\caption{
Ordering example code and execution. Each column represents a step of the global clock.
Each of the first four rows corresponds to a different static memory access. Values of thread local variables are shown in curly braces. The next two rows show the values of the shared variables.
Threads may arbitrarily stall but must maintain their relative ordering. This is shown by the stall of threads \(T1\) and \(T2\) at time 2.
}
\label{OrderedExecExample}
\begin{tikzpicture}[
font=\scriptsize,
t0graphnode/.style={rectangle, draw=black, fill=red!20, very thick, text height=0.25em, text width=2.25cm},
t1graphnode/.style={rectangle, draw=black, fill=blue!20, very thick, text height=0.25em, text width=2.25cm},
t2graphnode/.style={rectangle, draw=black, fill=green!20, very thick, text height=0.25em, text width=2.25cm},
node distance=0.25cm and 1.0cm
]

\node[t0graphnode]      (t0_0)                       {\makebox[0pt][l]{T0}\makebox[\textwidth][c]{\texttt{a = x;}}};
\node[t0graphnode]      (t0_1)       [below=of t0_0] {\makebox[0pt][l]{T0}\makebox[\textwidth][c]{\texttt{b = y;}}};
\node[t0graphnode]      (t0_2)       [below=of t0_1] {\makebox[0pt][l]{T0}\makebox[\textwidth][c]{\texttt{x = a+b;}}};
\node[t0graphnode]      (t0_3)       [below=of t0_2] {\makebox[0pt][l]{T0}\makebox[\textwidth][c]{\texttt{y = a*2;}}};

\node[t1graphnode]      (t1_0)       [right=of t0_0] {\makebox[0pt][l]{T1}\makebox[\textwidth][c]{\texttt{a = x;}}};
\node[t1graphnode]      (t1_1)       [below=of t1_0] {\makebox[0pt][l]{T1}\makebox[\textwidth][c]{\texttt{b = y;}}};
\node[t1graphnode]      (t1_2)       [below=of t1_1] {\makebox[0pt][l]{T1}\makebox[\textwidth][c]{\texttt{x = a+b;}}};
\node[t1graphnode]      (t1_3)       [below=of t1_2] {\makebox[0pt][l]{T1}\makebox[\textwidth][c]{\texttt{y = a*2;}}};

\node[t2graphnode]      (t2_0)       [right=of t1_0] {\makebox[0pt][l]{T2}\makebox[\textwidth][c]{\texttt{a = x;}}};
\node[t2graphnode]      (t2_1)       [below=of t2_0] {\makebox[0pt][l]{T2}\makebox[\textwidth][c]{\texttt{b = y;}}};
\node[t2graphnode]      (t2_2)       [below=of t2_1] {\makebox[0pt][l]{T2}\makebox[\textwidth][c]{\texttt{x = a+b;}}};
\node[t2graphnode]      (t2_3)       [below=of t2_2] {\makebox[0pt][l]{T2}\makebox[\textwidth][c]{\texttt{y = a*2;}}};

\draw[thick,->] (t0_0.south) -- (t0_1.north);
\draw[thick,->] (t0_1.south) -- (t0_2.north);
\draw[thick,->] (t0_2.south) -- (t0_3.north);

\draw[thick,->] (t1_0.south) -- (t1_1.north);
\draw[thick,->] (t1_1.south) -- (t1_2.north);
\draw[thick,->] (t1_2.south) -- (t1_3.north);

\draw[thick,->] (t2_0.south) -- (t2_1.north);
\draw[thick,->] (t2_1.south) -- (t2_2.north);
\draw[thick,->] (t2_2.south) -- (t2_3.north);

\draw[thick,dashed,->] (t0_0.east) -- (t1_0.west);
\draw[thick,dashed,->] (t0_1.east) -- (t1_1.west);
\draw[thick,dashed,->] (t0_2.east) -- (t1_2.west);
\draw[thick,dashed,->] (t0_3.east) -- (t1_3.west);

\draw[thick,dashed,->] (t1_0.east) -- (t2_0.west);
\draw[thick,dashed,->] (t1_1.east) -- (t2_1.west);
\draw[thick,dashed,->] (t1_2.east) -- (t2_2.west);
\draw[thick,dashed,->] (t1_3.east) -- (t2_3.west);
\end{tikzpicture}
\caption{Dependencies in ordering example.  The statements executed by a particular thread are stacked on top of each other.  Solid vertical dependence arcs represent the requirement that a particular thread must perform all memory accesses in program order.  Dashed horizontal dependence arcs represent inter-thread constraints imposed by \consistname{}.}
\label{OrderedExecConstraints}
\end{figure*}

A simple illustration of \consistname{} is shown in Figure \ref{OrderedExecExample}. This example contains four statements. The variables \texttt{x} and \texttt{y} are shared among threads, whereas \texttt{a} and \texttt{b} are thread-local. Each statement contains exactly one access to a shared variable. Figure \ref{OrderedExecExample} shows a valid execution of the ordering example by three threads (\(T0\), \(T1\), and \(T2\)). \(T0\) executes ahead of \(T1\), and \(T1\) executes ahead of \(T2\). A given thread executes statements from top to bottom (note that threads are not required to advance on each step of global clock, \eg{} time 2). Figure \ref{OrderedExecConstraints} shows dependencies between the four statements executed by the three threads. 

\subsection{Scheduling Constraints and Synchronization}\label{schedul_const}
Programmers can further limit the set of allowed executions of concurrent code using two language mechanisms: scheduling constraints and inter-thread synchronization.

Static scheduling constraints can be used to limit concurrency for a section of code, similar to how atomic read-modify-write operations limit the set of allowed executions in other consistency models \cite{hill-2020}. Scheduling constraints are expressed using the \texttt{[[schedule(N)]]} attribute, which applies the following restrictions to a block of code:

\begin{itemize}
    \item No more than \texttt{N} threads will be executing statements within the block at a time.
    \item All writes to shared state within the block will be scheduled for the same cycle (typically the last cycle in which a thread executes statements contained in the block).
    \footnote{When multiple writes are made to a shared variable within a block, intermediate values are local to a given thread.  At the end of the \texttt{[[schedule(N)]]} block, the final value is written to the shared variable.}
    \item If \texttt{N=1}, then all reads from shared state within the block for a given thread will be scheduled for the same cycle (not necessarily the same cycle as the writes).
\end{itemize}

The most commonly used \texttt{[[schedule(N)]]} is when \texttt{N=1}, and \langname{} defines a convenient alias for this case: \texttt{atomic}.  This is often used to protect a critical section.

The \texttt{[[thread\_rate(N)]]} attribute can be used to specify the initiation interval of a function, limiting the rate at which threads can enter the function.

Scheduling constraints are used to implement the statically scheduled variant of \texttt{CountIf \allowbreak Histogram} (option 1). Figure \ref{StaticScheduleCIH} shows the code\footnote{When reading implementations of the running example, note that the variables \texttt{feature}, \texttt{weight}, \texttt{hist}, and \texttt{locks} are shared between threads. All other variables are thread-local.}. The implementation makes use of \texttt{pipelined\_for}\footnote{\texttt{pipelined\_for} is a thin wrapper around a call to a batched function (see Section \ref{BatchedFunctions}).  It is named \texttt{pipelined\_for} rather than \texttt{parallel\_for} to express the notion that threads begin execution of the loop body in a well-defined order.}, a library function that executes a user-specified function \texttt{N} times, each with a separate thread.  The \texttt{atomic} scheduling constraint ensures that only one thread can perform the read-modify-write sequence at a time. The \texttt{[[thread\_rate(L)]]} attribute sets the initiation interval of \texttt{body} to \(L\).

Inter-thread synchronization in \langname{} can be expressed using common synchronization objects like mutexes, semaphores, or read-write locks, which are provided by the standard library. Most are built on top of a single language primitive, \texttt{wait\_for}, which combines the \texttt{atomic} scheduling constraint with a loop. A statement \texttt{wait\_for(expr)} evaluates \texttt{expr} for a single thread until \texttt{expr} evaluates to true. Accesses to shared variables within \texttt{expr} occur at the same time. \texttt{wait\_for} maintains thread ordering, thus exhibiting head-of-line blocking.

Figure \ref{DynamicallyScheduledCIH} shows \langname{} code implementing dynamic scheduling (option 2). The shared variable \texttt{locks} implements an array of mutexes\footnote{An idiomatic implementation would use a library synchronization object rather than using \texttt{wait\_for} directly, but the low-level version is shown here for clarity.}. \texttt{wait\_for} is used to stall a thread if there are in-flight accesses to the same histogram bucket. This code passes a lambda to \texttt{pipelined\_for} rather than using a named function like the previous example. 

Figure \ref{CSlowCIH} demonstrates how to use scheduling constraints to implement a version of \texttt{CountIf Histogram} that operates on \(L\) separate histograms, all stored in the \texttt{hist} array (option 3). The \texttt{[[schedule(L)]]} constraint ensures that no more than \(L\) threads can be executing the read-modify-write sequence concurrently. Using \texttt{offset} ensures that each of these threads operates on a different intermediate histogram, eliminating the hazard. A post-processing step is added after the first loop, computing the final result. Here, a sequential \texttt{for} loop (discussed in more detail in Section \ref{SequentialLoops}) is nested inside the function called by \texttt{pipelined\_for}. At any moment in time, there can be many threads executing the body of the \texttt{for} loop, each with a distinct value of \texttt{m}.

\begin{figure*}[t]
  \belowcaptionskip = 0pt
  \begin{minipage}[t]{.51\linewidth}
    \begin{lstlisting}[language=Ibis]
void static_count_if() {
  [[thread_rate(L)]] void body(uint32 i) {
    int32 m = feature[i];
    float32 wt = weight[i];
    if (m > THRESHOLD){
      atomic {
        float32 x = hist[m];
        hist[m] = x + wt;
      }
    }
  }

  pipelined_for(N, body);
}
    \end{lstlisting}
    \caption{Statically scheduled \texttt{CountIf Histogram}.}
    \label{StaticScheduleCIH}
  \end{minipage}%
  \hfill%
  \begin{minipage}[t]{.43\linewidth}
    \begin{lstlisting}[language=Ibis]
bool[SIZE] locks = {};
inline bool lock(int32 m) {
  bool result = !locks[m];
  if (result) { locks[m] = true; }
  return result;
}
void dynamic_count_if() {
  pipelined_for(N, [](uint32 i) {
    int32 m = feature[i];
    float32 wt = weight[i];
    if (m > THRESHOLD) {
      wait_for(lock(m));
      float32 x = hist[m];
      hist[m] = x + wt;
      locks[m] = false;
    }
  });
}
    \end{lstlisting}
    \caption{Dynamically scheduled \texttt{CountIf Histogram}.}
    \label{DynamicallyScheduledCIH}
  \end{minipage}
\end{figure*}
\begin{figure*}[t]
  \belowcaptionskip = 0pt
  \begin{minipage}[t]{.51\linewidth}
    \begin{lstlisting}[language=Ibis]
void replicated_count_if() {
  pipelined_for(N, [](uint32 i) {
    int32 m = feature[i];
    float32 wt = weight[i];
    if (m > THRESHOLD) {
      uint32 offset = (i % L) * SIZE;
      [[schedule(L)]] {
        float32 x = hist[m + offset];
        hist[m + offset] = x + wt;
      }
    }
  });
  pipelined_for(SIZE, [](uint32 m) {
    float32 sum = 0.0;
    for (const auto l : L) {
      sum += hist[m + (l * SIZE)];
    }
    hist[m] = sum;
  });
}
    \end{lstlisting}
    \caption{Replicated \texttt{CountIf Histogram}.}
    \label{CSlowCIH}
  \end{minipage}%
  \hfill%
  \begin{minipage}[t]{.43\linewidth}
    \begin{lstlisting}[language=C++]
if (c) {
  shared_var = 1;
} else {
  x = shared_var;
}
    \end{lstlisting}
    \caption{Ordering of conditional shared variable access.}
    \label{OrderingOfConditionalSVA}
  \end{minipage}
\end{figure*}

\subsection{Thread Ordering with Control Flow} 
\label{sec:ControlFlow}

In this section, we describe how thread ordering is affected by \langname{} control flow constructs.

\subsubsection{Function Calls}
If thread \(T0\) reaches a function call ahead of thread \(T1\), then thread \(T0\) will begin executing the body of the called function ahead of \(T1\). 

\subsubsection{Asynchronous Functions} 
\langname{} supports the concept of an asynchronous function call, in which a new thread executes the called function completely decoupled from the calling thread. This is the only case when total ordering among threads is not defined. If thread \(T0\) reaches an asynchronous function call before thread \(T1\), a new thread \(T0'\) will start executing the body of the function with parameters from \(T0\), ahead of another new thread \(T1'\) with parameters from thread \(T1\). No ordering is defined between threads \(T0\) and \(T0'\) nor \(T1\) and \(T1'\).  Threads \(T0\) and \(T1\) continue executing without waiting for the function to return and their relative order is preserved. Asynchronous functions are building blocks which are commonly used in the implementation of control flow libraries like fork-join and rendezvous \cite{ada-1983}.

\subsubsection{Batched Functions}
\label{BatchedFunctions}
\langname{} supports structured concurrency with batched functions. When a thread calls a batched function, the function body is executed \(N\) times by a group of \(N\) threads, where \(N\) is specified by the first argument of the function call. The calling thread blocks until all nested threads finish execution. The value of the first parameter within the function body specifies an index of the current thread within the group, with the first thread starting at index 0. The remaining parameter values are set to argument values from the calling thread. If thread \(T0\) reaches a call to a batched function ahead of thread \(T1\), then all nested threads associated with \(T0\) begin execution of the function ahead of nested threads associated with thread \(T1\).

\langname{} models structured concurrency as calls to batched functions to preserve the strict separation of thread-local states. The threads which execute a batched function do not have access to local state of the calling thread. Any data from the calling thread must be explicitly passed via parameters to the batched function.  The standard library functions like \texttt{pipelined\_for} provide convenient wrappers around batched functions that can be used like loops, with lexical closures used to explicitly pass values from thread-local state when needed.

\subsubsection{Loops}
\label{SequentialLoops}
Although the best performance is usually achieved by expressing loops using library functions like the aforementioned \texttt{pipelined\_for}, \langname{} also supports sequential \texttt{for} and \texttt{do-while} loops. Sequential loops are executed by existing threads just like any other statements. Each thread reaching a loop will execute all iterations of the loop before continuing to statements following the loop\footnote{Multiple threads may be executing the loop concurrently.}. If thread \(T0\) reaches a loop ahead of thread \(T1\), then thread \(T0\) will start executing the first iteration of the loop ahead of \(T1\). The threads start executing subsequent iterations in the order in which they reach end of the loop body on the previous iteration.
Threads executing subsequent iterations are also given priority over new threads wishing to enter the loop.

\langname{} supports two different behaviors for threads exiting a loop.  Ordered loops always preserve thread order. Logically, there is a barrier which ensures threads begin executing the statement following the loop in the order in which they entered the loop (even if the threads executed a different number of loop iterations).  With unordered loops\footnote{An attribute \texttt{[[unordered]]} is used to mark a loop as unordered.}, threads begin executing the statement following the loop in the order in which they reach the end of the loop (\ie{} finish executing the last iteration). 

\subsubsection{Conditionals}
\label{subsec:conditionals}
Conditionals do not affect thread ordering.  When a thread reaches a conditional, it executes statements in both branches, predicating side-effects based on the condition.  The thread-ordering restriction of \consistname{} applies to all shared variable accesses, including the accesses which are disabled by the predicate.  For example, in Figure \ref{OrderingOfConditionalSVA} if thread \(T0\) (with \texttt{c=true}) reaches the conditional ahead of thread \(T1\) (with \texttt{c=false}), then the read executed by thread \(T1\) will execute after the write executed by thread \(T0\).  However, if the conditional in Figure \ref{OrderingOfConditionalSVA} were reversed such that the read of \texttt{shared\_var} appeared before the write, there is no guarantee that \(T0\) (executing the write) would update \texttt{shared\_var} before \(T1\) read it.

\subsection{Parallelism}
\label{Parallelism}
Pipeline parallelism is the primary way of achieving parallelism in \langname{}.  Here we describe language constructs to replicate design elements if additional parallelism is needed.

\subsubsection{Functions}
Function calls can be inlined, effectively replicating the body of the function at the call site.  Each replica contains replicas of all control flow constructs, scheduling constraints, and \texttt{static} local variables from the inlined function.  

\subsubsection{Classes}
As in other object-oriented languages, classes in \langname{} contain both code and data and support information hiding.  There is no dynamic allocation of objects in \langname{}, and thus the number of objects is known at compile time.  Each instance of a class results in the replication of member variables and non-inline methods of that class.

\subsubsection{Compile Time Control}
\texttt{static for} loops execute at compile time.  Each iteration of a \texttt{static for} loop defines unique instances for each control flow construct and memory access in the loop body.  In some cases, it is more natural to express compile time iteration with recursion rather than loops. Figure \ref{MapReduce} shows examples of both static loops and static recursion.

\begin{figure}[t]
  \belowcaptionskip = 0pt
  \begin{minipage}[t]{.55\linewidth}
    \begin{lstlisting}[language=Ibis]
template <typename T, auto N>
inline auto map((T) -> auto f, T[N] x) {
  using result_t = decltype(f(x[0]));
  result_t[N] r;
  static for(const auto i : N) {
    r[i] = f(x[i]);
  }
  return r;
}
template <typename T, auto N>
inline T reduce((T, T) -> T f, T[N] x) {
  static if (N == 1) {
    return x[0];
  } else {
    const auto NewN = (N + 1) / 2;
    T[NewN] new_array;
    static for(const auto i : N/2) {
      new_array[i] = f(x[2*i], x[2*i + 1]);
    }
    static if ((N % 2) == 1) {
      new_array[NewN - 1] = x[N - 1];
    }
    return reduce(f, new_array);
  }
}
template <typename T, auto N, typename R>
inline R map_reduce((T) -> R map_fn,
                    (R, R) -> R reduce_fn,
                    T[N] x) {
  return reduce(reduce_fn, map(map_fn, x));
}
    \end{lstlisting}%
    \caption{Map-reduce implementation.}%
    \label{MapReduce}%
  \end{minipage}%
\end{figure}%

\subsection{Code Composability and Reuse}
Composable and reusable code is taken for granted in software programming, but has been historically elusive for hardware designers, especially at smaller granularities of individual data structures, algorithms and design patterns. 

\langname{} facilitates building composable components by including various language constructs for creating abstractions without incurring extra hardware costs. The objective is to eliminate any justification for the ad-hoc coding style that relies on copy-and-paste as a code reuse mechanism, a practice all too common in hardware design. To achieve this, we draw from established programming language design concepts, such as first-class functions \cite{cisp-1985}, lambdas \cite{lambda-imperative-1976, lambda-declarative-1976}, lexical closures, object orientation, and modularity. We defined their semantics in the language to specifically adhere to the zero-cost abstraction principle \cite{design-cpp-1994}.

In \langname{}, functions are first-class values and can be stored in data structures, passed as function arguments, or returned as function results. The support for higher-order functions is complemented by the inclusion of lambdas (anonymous functions) and lexical closures. A key element in achieving the zero-cost abstraction principle with respect to first-class functions is the static resolution of all function calls at compile time. \langname{} does not employ the concept of function pointers nor dynamic dispatch function calls. Similarly lexical closures are implemented in a way that does not incur any overhead compared to regular function arguments.

Functions can be generic in terms of their parameters and return types. The static type system ensures compile-time type correctness while type inference frees the programmer from adding tedious type annotations. A strong static type system helps eliminate whole categories of bugs, which can be especially important in hardware designs.

\langname{} comes with an extensive standard library that includes common data structures, algorithms and reusable implementations of idiomatic coding patterns. As an illustrative example, Figure \ref{MapReduce} shows the implementation of a map-reduce algorithm as a generic higher-order function. It takes user-defined map and reduce functions as arguments, and uses compile-time recursion to implement a reduction tree which has a depth that depends on input size. The function is generic not only in terms of the size and type of its input but also logic depth and pipeline latency required for the map and reduce functions. 

The library also contain larger Intellectual Property (IP) blocks, such as RISC-V processor, IEEE-754 floating point arithmetic library, compression codecs, cryptographic algorithms, and many others.

\section{Advanced \langname{} Example - Speculative CountIf Histogram}
\label{sec:complex}
The implementation of the speculative version of \texttt{CountIf Histogram} (option 4) is more complex than the other examples. It takes advantage of various features in \langname{} for code composability and reuse.

We first implement a generic speculative loop, shown in Figure \ref{SpeculativePipelinedFor}. Similar to the \texttt{pipelined\_for}, the iterations of \texttt{spec\_pipelined\_for} are concurrently executed by multiple threads. Loop iteration is governed by two user-provided functions: \texttt{body}, which speculatively generates results for a given iteration index, and \texttt{try\_commit}, which commits the results after verifying that they have not been invalidated by another iteration running concurrently. In the event a mis-speculation is detected, the in-flight threads are drained without committing their calculated results. Subsequently, the loop restarts execution on the iteration that failed to commit.

When a thread executes the \texttt{atomic} block on line 13, it chooses a speculative loop iteration index \texttt{i} that it will execute.  The first thread to enter this atomic block after a mis-speculation occurs will see \texttt{top\_ok} is \texttt{false}.  This thread is called a "flush thread".  A flush thread will set \texttt{i} to the lowest iteration index that has not yet been committed (flush threads do not speculate).  The following threads will speculate, incrementing loop iteration index values.

After choosing an iteration index, threads execute \texttt{body}, which can be arbitrarily complex (\eg{} contain deeply nested control flow).  Next, threads execute the atomic block on line 24.  In this block, a thread will determine if it should commit its results.  Results are committed if there is not an unresolved mis-speculation and \texttt{try\_commit} returns \texttt{true}.  If \texttt{try\_commit} returns \texttt{false}, then a new mis-speculation has been detected.  That mis-speculation will be resolved when a flush thread reaches the atomic block on line 24.

\begin{figure*}[h!]
  \belowcaptionskip = 10pt
  \begin{minipage}[t]{.7\linewidth}
    \begin{lstlisting}[language=Ibis]
template <typename C, typename I, typename B>
inline void spec_pipelined_for(
  C count, 
  (I)->B body, 
  (B)->bool try_commit) {
  pipelined_do<uint8>([count, body, try_commit](uint8 _) {
    static bool top_ok = true;
    static bool bottom_ok = true;
    static I speculative_iteration = 0;
    static C completed_iterations = 0;
    I i;
    bool is_flush_thread = false;
    atomic {
      if (!top_ok){
        speculative_iteration = completed_iterations;
        is_flush_thread = true;
        top_ok = true;
      }
      i = speculative_iteration;
      speculative_iteration++;
    }
    B body_result = body(i);
    bool loop_is_done = false;
    atomic {
      loop_is_done = completed_iterations == count;
      if (is_flush_thread) { bottom_ok = true; }
      if (bottom_ok && !loop_is_done) {
        bool commit_result = try_commit(body_result);
        if (!commit_result) {
          bottom_ok = false;
          top_ok = false;
        } else { completed_iterations++; }
      }
    }
    return !loop_is_done;
  });
}
    \end{lstlisting}
    \caption{Speculative loop.}
    \label{SpeculativePipelinedFor}
  \end{minipage}
  \begin{minipage}[t]{.7\linewidth}
    \begin{lstlisting}[language=Ibis]
void spec_count_if() {
  struct context{ int32 m; float32 prev; float32 sum; };
  spec_pipelined_for(N, [](uint32 i) {
    int32 m = feature[i];
    float32 wt = weight[i];
    float32 prev = hist[m];
    context c = {
      .m = m, 
      .prev = prev, 
      .sum = m > THRESHOLD ? prev + wt : prev
    };
    return c;
  }, [](context c) {
    bool result = false;
    float32 prev = hist[c.m];
    if (eq(prev, c.prev)) {
      result = true;
      hist[c.m] = c.sum;
    }
    return result;
  });
}
    \end{lstlisting}
    \caption{Speculative \texttt{CountIf Histogram}.}
    \label{SpeculativeCountIf}
  \end{minipage}
\end{figure*}

Figure \ref{SpeculativeCountIf} implements \texttt{CountIf Histogram} with a call to \texttt{spec\_pipelined\_for}.  The \texttt{body} function reads input data and speculatively reads a histogram value and computes a new result.  \texttt{try\_commit} compares a current histogram value with the prior speculative read.  If these values match then the loop iteration is allowed to complete.

Implementing a computation like this would be difficult with traditional HLS because an HLS compiler would see false dependencies:
\begin{itemize}
    \item Read-after-write and write-after-write hazards on \texttt{top\_ok}.
    \item Read-after-write hazards on \texttt{hist[m]}.
\end{itemize}

The \langname{} compiler does not infer these dependencies because \texttt{top\_ok} and \texttt{hist} are shared variables, and it is up to the programmer to handle inter-thread hazards.

Implementing a computation like this with traditional (\ie{} unordered) threads would be difficult because threads could overtake each other while executing \texttt{body}.  Threads would either need to speculate that they would remain in order (which would frequently not be true), or a costly synchronization mechanism would be needed to restore order after executing \texttt{body}.

\section{Translation to Hardware}
\label{sec:translation}
In this section, we describe a mapping from \langname{} source code to hardware circuits suitable for FPGA or ASIC implementation.  The overall compiler flow is discussed, followed by the implementation of specific constructs.

\subsection{Compiler Overview}
After parsing, template instantiation, and type-checking, the \langname{} compiler converts the abstract syntax tree (AST) into a typical imperative Intermediate Representation (IR).  Each function in the IR is represented as a control flow graph.  

If-conversion \cite{Ferrante-1987} is applied to flatten all conditional statements, meaning that they will not break up the control flow graph.  This approach is a straightforward method to implement the semantics described in Section \ref{subsec:conditionals}.  Functions with a single call site that do not have additional attributes (\eg{} asynchronous functions) are automatically inlined.  Traditional peephole \cite{peephole-1965} and data-flow analysis \cite{dfa-1973} optimizations are applied on the IR.  The primary purpose of these optimizations is to reduce hardware resource usage.

Next, the operations in each basic block are scheduled into a pipeline and imperative control is lowered to hardware flow control.
Finally, \langname{} utilizes CIRCT \cite{CIRCT} to generate SystemVerilog output.

\subsection{Control Flow Graph and Pipelines}
There is a direct correspondence between the control-flow graph describing the source design and the generated hardware.  Figure \ref{rci-cfg} shows such a control-flow graph for \texttt{replicated\_count\_if}\footnote{A \texttt{return} statement has been added for clarity.}.  Each basic block in the control-flow graph corresponds to a pipeline (comprising one or more pipeline stages) in the generated hardware and each edge in the control-flow graph corresponds to communication between pipelines, usually through FIFOs (\ie{} first-in-first-out queues of bounded size).  A key property of this translation is that it can be described as a recursive composition of translations, specific to each control flow construct.  For example, the translation of a \texttt{for} loop to hardware can be described independent of the context in which that loop appears.

The computation performed by a single pipeline stage can potentially execute in parallel with the computation performed by all other pipeline stages.  The amount of concurrency (\ie{} active threads) in a design determines how much of this potential is utilized.  At any moment in time, each thread is either assigned to a single pipeline stage or is queued in a FIFO\footnote{The values of live local variables are stored in pipeline registers which are input to the pipeline stage where the thread resides, or inside the FIFO where the thread resides.}.  Two threads cannot be assigned to the same pipeline stage at the same time.  The values of local variables associated with a thread are stored in pipeline registers and FIFOs.  This state flows through the generated hardware (within and between basic blocks), accompanying their threads as they flow through the control-flow graph of the source design.

\begin{figure*}[t]
\begin{subfigure}[h]{0.5\linewidth}
\begin{lstlisting}[language=Ibis]
void replicated_count_if() {
  // Basic Block 0
  pipelined_for(N, [](uint32 i) {
    // Basic Block 1
    int32 m = feature[i];
    float32 wt = weight[i];
    if (m > THRESHOLD) {
      uint32 offset = (i % L) * SIZE;
      [[schedule(L)]] {
        float32 x = hist[m + offset];
        hist[m + offset] = x + wt;
      }
    }
  });
  // Basic Block 2
  pipelined_for(SIZE, [](uint32 m) {
    // Basic Block 3
    float32 sum = 0.0;
    for (const auto l : L) {
      // Basic Block 4
      sum += hist[m + (l * SIZE)];
    }
    // Basic Block 5
    hist[m] = sum;
  });
  // Basic Block 6
  return;
}
\end{lstlisting}
\end{subfigure}
\hfill
\begin{subfigure}[h]{0.4\linewidth}
\begin{tikzpicture}[
every node/.style={font=\scriptsize},
stage/.style={rectangle, minimum width={12em}, draw=black, align=left, text width=45mm},
roundedrect/.style={rectangle, rounded corners, draw=black, very thick, minimum size=5mm},
fsm/.style={circle, draw=black, thick, align=center, minimum size=4mm},
]

\node[stage]                (bb0)                        {\texttt{\HS pipelined\_for(N, ...)}};
\node[stage]                (bb1)    [below=4mm of bb0]  {\texttt{\HS int32 m = feature[i];}\\ 
                                                          \texttt{\HS float32 wt = weight[i];}\\ 
                                                          \texttt{\HS if (m > THRESHOLD) \{}\\
                                                          \texttt{\HS\HS\HS uint32 offset = (i \% L) * SIZE;}\\
                                                          \texttt{\HS\HS\HS [[schedule(L)]] \{}\\
                                                          \texttt{\HS\HS\HS\HS\HS float32 x = hist[m + offset];}\\
                                                          \texttt{\HS\HS\HS\HS\HS hist[m + offset] = x + wt}\\
                                                          \texttt{\HS\HS\HS\}}\\ 
                                                          \texttt{\HS\}}}; 
\node[stage]                (bb2)    [below=4mm of bb1]   {\texttt{\HS pipelined\_for(SIZE, ...)}};
\node[stage]                (bb3)    [below=4mm of bb2]   {\texttt{\HS float32 sum = 0.0;}};
\node[stage]                (bb4)    [below=4mm of bb3]   {\texttt{\HS sum += hist[m + (l * SIZE)];}};
\node[stage]                (bb5)    [below=4mm of bb4]   {\texttt{\HS hist[m] = sum;}};
\node[stage]                (bb6)    [below=4mm of bb5]   {\texttt{\HS return;}};

\draw[thick,->] (bb0.south)     -- (bb1.north);
\draw[thick,->] (bb1.south)     -- (bb2.north);
\draw[thick,->] (bb2.south)     -- (bb3.north);
\draw[thick,->] (bb3.south)     -- (bb4.north);
\draw[thick,->] (bb4.south)     -- (bb5.north);
\draw[thick,->] (bb4.310)       -- ++(0,-0.2) |- ++(2.5,0) |- ++(0,0.9) |- ++(-2.46,0) -- (bb4.45);
\draw[thick,->] (bb5.south)     -- (bb6.north);

\end{tikzpicture}
\end{subfigure}
\caption{Control-flow graph of \texttt{replicated\_count\_if}.}
\label{rci-cfg}
  \begin{subfigure}[h]{0.5\linewidth}
  \begin{lstlisting}[language=Ibis]
void replicated_count_if() {
  <body1>
}
  \end{lstlisting}
  \end{subfigure}
  \hfill
  \begin{subfigure}[h]{0.4\linewidth}
  \begin{tikzpicture}[
  every node/.style={font=\scriptsize},
  thickrect/.style={rectangle, draw=black, very thick, minimum size=5mm},
  roundedrect/.style={rectangle, rounded corners, draw=black, very thick, minimum size=5mm},
  node distance=0.3cm and 0.5cm
  ]
  
  \node[queue={Argument\\FIFO},minimum height=0.5cm]    (args)                  {};
  \node[roundedrect]              (body) [below=of args]  {body1};
  \node[queue={Return\\FIFO},minimum height=0.5cm]      (rets) [below=of body]  {};
  \node[queue={},label={[label distance=-0.15cm]right:{\begin{tabular}{c}Context\\FIFO\end{tabular}}},minimum height=0.5cm] (context) [right=of body] {};
  
  \draw[thick,->] (args.south) -- (body.north);
  \draw[thick,->] (body.south) -- (rets.north);
  \draw[thick,->] (context.north |- current bounding box.north) -- (context.north);
  \draw[thick,->] (context.south) -- (context.south |- current bounding box.south);
  
  \end{tikzpicture}
  \end{subfigure}
  \caption{Hardware realization of a function.}
  \label{translation-fn}
\end{figure*}

\subsection{Translation Example}
\subsubsection{Function}
Figure \ref{translation-fn} illustrates the hardware realization of a non-inlined function using a fixed-latency pipeline and latency-insensitive FIFOs.  The caller enqueues thread-local data into a set of two input FIFOs: argument and context.  The argument FIFO holds values that are inputs to the entry basic block of the function. A hidden \texttt{call-site-id} argument is also enqueued into the argument FIFO, indicating the source of the function call.  The context FIFO holds the values of thread-local variables that are live at the time of the function call.  When the function returns, it places return values in a return FIFO.  Function returns are implemented with dynamic dispatch and the \texttt{call-site-id} parameter determines which return FIFO receives the values.  The basic block following each function call site combines the returned values with the stored thread-local variables from the context FIFO to continue execution.  Note that the input and output FIFOs are still logically present in the design even if there are no arguments or return values, because they count the number of threads contained in each FIFO, tracking function entry and return.  \texttt{body1} represents the body of the function, described in Figure \ref{translation-batched}.

\subsubsection{Batched Function}
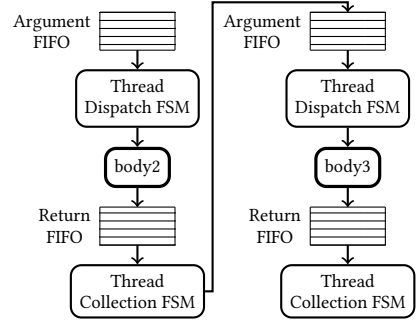
\begin{figure*}[t]
  \belowcaptionskip = 0pt
  \begin{subfigure}[h]{0.5\linewidth}
    \begin{lstlisting}[language=Ibis]
void replicated_count_if() {
  pipelined_for(N, [](uint32 i) {
    <body2>
  });

  pipelined_for(SIZE, [](uint32 m) {
    <body3>
  });
}
    \end{lstlisting}
  \end{subfigure}
  \hfill
  \begin{subfigure}[h]{0.45\linewidth}
    \begin{tikzpicture}[
    every node/.style={font=\scriptsize},
    thickrect/.style={rectangle, draw=black, very thick, minimum size=5mm},
    roundedrect/.style={rectangle, rounded corners, draw=black, very thick, minimum size=5mm, align=center},
    fsm/.style={rectangle, rounded corners, draw=black, thick, minimum size=5mm, align=center},
    node distance=0.25cm and 1.0cm
    ]
    
    \node[queue={Argument\\FIFO},minimum height=0.5cm]    (mlargs)    {};
    \node[fsm]                      (mlfsm1)   [below=of mlargs]   {Thread\\Dispatch FSM};
    \node[roundedrect]              (body2)    [below=of mlfsm1]   {body2};
    \node[queue={Return\\FIFO},minimum height=0.5cm]      (mlrets)   [below=of body2]    {};
    \node[fsm]                      (mlfsm2)   [below=of mlrets]   {Thread\\Collection FSM};
    
    \draw[thick,->] (mlargs.south)  -- (mlfsm1.north);
    \draw[thick,->] (mlfsm1.south)  -- (body2.north);
    \draw[thick,->] (body2.south)   -- (mlrets.north);
    \draw[thick,->] (mlrets.south)  -- (mlfsm2.north);
    
    \node[queue={Argument\\FIFO},minimum height=0.5cm]    (ppargs)   [right=18mm of mlargs]{};
    \node[fsm]                      (ppfsm1)   [below=of ppargs]    {Thread\\Dispatch FSM};
    \node[roundedrect]              (body2)    [below=of ppfsm1]    {body3};
    \node[queue={Return\\FIFO},minimum height=0.5cm]      (pprets)   [below=of body2]     {};
    \node[fsm]                      (ppfsm2)   [below=of pprets]    {Thread\\Collection FSM};
    
    \draw[thick,->] (mlfsm2.east)   -- ++(0.1,0) |- ++(0,3.8) -| (ppargs.north);
    \draw[thick,->] (ppargs.south)  -- (ppfsm1.north);
    \draw[thick,->] (ppfsm1.south)  -- (body2.north);
    \draw[thick,->] (body2.south)   -- (pprets.north);
    \draw[thick,->] (pprets.south)  -- (ppfsm2.north);
    
    \end{tikzpicture}
  \end{subfigure}%
  \caption{Hardware realization of a call to a batched function (body1). Context FIFOs omitted for clarity.}
  \label{translation-batched}
\end{figure*}

Figure \ref{translation-batched} illustrates the hardware implementation of calls to batched functions.  The body of each function is wrapped with two finite state machines (FSMs).  The thread dispatch FSM dispatches a group of threads into the function body (at a maximum rate of one per clock cycle).  The thread collection FSM waits for all threads in a group to finish execution of the function before enqueuing results into the return FIFO.  Note that the generated hardware supports nested pipeline parallelism.

\begin{figure*}[t]
  \belowcaptionskip = 0pt
  \begin{subfigure}[h]{0.5\linewidth}
    \begin{lstlisting}[language=Ibis]
pipelined_for(N, [](uint32 i) {
  int32 m = feature[i];
  float32 wt = weight[i];
  if (m > THRESHOLD) {
    uint32 offset = (i % L) * SIZE;
    [[schedule(L)]] {
      float32 x = hist[m + offset];
      hist[m + offset] = x + wt;
    }
  }
});
    \end{lstlisting}
  \end{subfigure}
  \hfill
  \begin{subfigure}[h]{0.4\linewidth}
    \begin{tikzpicture}[
    every node/.style={font=\scriptsize},
    stage/.style={rectangle, minimum width={12em}, draw=black, align=left, text width=45mm},
    roundedrect/.style={rectangle, rounded corners, draw=black, very thick, minimum size=5mm},
    fsm/.style={circle, draw=black, thick, align=center, minimum size=4mm},
    ]
    
    \node[stage] (s0)                       {\texttt{\HS int32 m = feature[i]};\\ \texttt{\HS float32 wt = weight[i];}};
    \node[stage] (s1) [below=2mm of s0]     {\texttt{\HS bool p = m > THRESHOLD;}\\ \texttt{\HS uint32 offset = (i \% L) * SIZE;}};
    \node[stage] (s2) [below=2mm of s1]     {\texttt{\HS [p] float32 x = hist[m + offset];}};
    \node[stage] (s3) [below=2mm of s2]     {\texttt{\HS [p] float32 t0 = x + wt;}};
    \node[stage] (s4) [below=2mm of s3]     {\texttt{\HS [p] hist[m + offset] = t0;}};
    
    \draw[thick,->] (s0.south)  -- (s1.north);
    \draw[thick,->] (s1.south)  -- (s2.north);
    \draw[thick,->] (s2.south)  -- (s3.north);
    \draw[thick,->] (s3.south)  -- (s4.north);
    
    \end{tikzpicture}
  \end{subfigure}%
  \caption{Hardware realization of a basic block (body2).}
  \label{translation-body2}
\end{figure*}
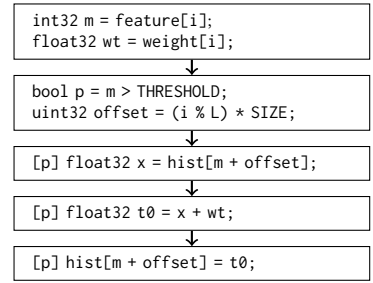
\begin{figure*}[t]
  \belowcaptionskip = 0pt
  \begin{subfigure}[h]{0.5\linewidth}
    \begin{lstlisting}[language=Ibis]
float32 sum = 0.0;
for (const auto l : L) {
  sum += hist[m + (l * SIZE)];
}
hist[m] = sum;
    \end{lstlisting}
  \end{subfigure}
  \hfill
  \begin{subfigure}[h]{0.4\linewidth}
    \begin{tikzpicture}[
    every node/.style={font=\scriptsize},
    stage/.style={rectangle, minimum width={12em}, draw=black, align=left, text width=45mm},
    roundedrect/.style={rectangle, rounded corners, draw=black, very thick, minimum size=5mm},
    fsm/.style={circle, draw=black, thick, align=center, minimum size=4mm},
    ]
    
    \node[stage]                (s0)                            {\texttt{\HS float32 sum = 0.0;}};
    \node[queue={Entry\\FIFO},minimum height=0.5cm]   (entry) [below=2mm of s0]       {};
    \node[stage]                (s1)    [below=2mm of entry]    {\texttt{\HS uint32 t0 = m + (l * SIZE);}};
    \node[stage]                (s2)    [below=2mm of s1]       {\texttt{\HS float32 t1 = hist[t0];}};
    \node[stage]                (s3)    [below=2mm of s2]       {\texttt{\HS sum += t1;}};
    \node[queue={Exit\\FIFO},minimum height=0.5cm]    (exit)  [below=2mm of s3]       {};
    \node[stage]                (s4)    [below=2mm of exit]     {\texttt{\HS hist[m] = sum;}};
    
    \draw[thick,->] (s0.south)     -- (entry.north);
    \draw[thick,->] (entry.south)  -- (s1.north);
    \draw[thick,->] (s1.south)     -- (s2.north);
    \draw[thick,->] (s2.south)     -- (s3.north);
    \draw[thick,->] (s3.south)     -- (exit.north);
    \draw[thick,->] (s3.310)       -- ++(0,-0.1) |- ++(2.7,0) |- ++(0,1.98) -| (s1.45);
    \draw[thick,->] (exit.south)   -- (s4.north);
    
    \end{tikzpicture}
  \end{subfigure}%
  \caption{Hardware realization of a loop (body3).}
  \label{translation-body3}
\end{figure*}
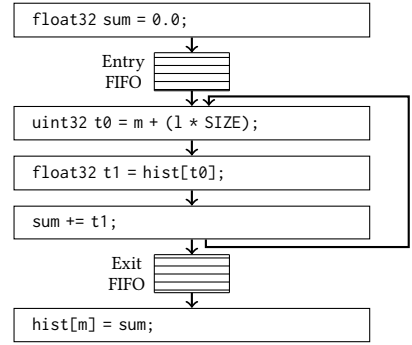
\subsubsection{Basic Block}
Figure \ref{translation-body2} illustrates how a basic block (\texttt{body2}) translates to a pipeline \footnote{Complex expressions have been decomposed into simpler ones and if-conversion has been applied to the \texttt{if} statement.}.  The operations which comprise a basic block are scheduled into pipeline stages.  A list scheduler is used to find a valid schedule with respect to the following constraints:
\begin{itemize}
    \item Intra-thread data dependencies are honored for accesses to thread-local variables.
    \item \consistname \:is preserved for accesses to shared variables.
    \item User-specified scheduling constraints (\texttt{[[schedule()]]}, \texttt{atomic}).
    \item Target logic depth per pipeline stage\footnote{The number of logic operations between registers.  This is specified as a compilation parameter outside of the code.  It can be swept to fine-tune the design, finding a balance between clock frequency and resource use.}.
\end{itemize}

Scheduling constraints do not explicitly appear in the scheduled pipeline, because they are satisfied by the schedule.  The requirements of \consistname{} are honored by ensuring that the static schedule of accesses to shared variables by a single thread occur in program order.  No special effort is needed to maintain inter-thread order because hardware pipelines maintain order by construction.  
The logic depth constraint is a soft constraint, used to assist in meeting a target operating frequency for the generated circuit, similar to scheduling in traditional HLS tools \cite{sdc-2006, sdc-13}. It will be violated if necessary to honor the other constraints.  It is assumed that the EDA tools consuming the output SystemVerilog will compensate by giving higher priority to cells along paths which violate the logic depth constraint when considering placement, routing, and in the case of an ASIC flow, logic cell threshold voltage and drive strength.

Beyond meeting the hard and soft constraints above, the current compiler does not yet include significant scheduling optimizations, a potential area for improvement.  For example, \citet{sdc-13} uses linear programming to minimize value lifetimes, which reduces the number of pipeline registers. 

\subsubsection{Sequential Loop}
Figure \ref{translation-body3} illustrates how a sequential loop \footnote{A single thread executes the loop sequentially, but multiple threads can execute the loop concurrently.} (\texttt{body3}) is realized.  The body of the loop is implemented as a pipeline.  Threads enter that pipeline either from the loop entry FIFO, or via a loopback path.  When a thread reaches the end of the loop, it is either sent back to the start of the loop body for another iteration or it is sent into the loop exit FIFO, which is read by the pipeline corresponding to the basic block after the loop.  Note that while in this case the loop body is a single basic block, in general the loop body can be an arbitrary control-flow graph.

\begin{figure*}[t]
  \begin{subfigure}[h]{0.5\linewidth}
      \begin{lstlisting}[language=Ibis]
void function_with_wait_for() {
<Statements before wait_for>
wait_for(expr);
<Statements after wait_for>
}
      \end{lstlisting}
  \end{subfigure}
  \hfill
  \begin{subfigure}[h]{0.4\linewidth}
      \begin{tikzpicture}[
          every node/.style={font=\scriptsize},
          thickrect/.style={rectangle, draw=black, very thick, minimum size=5mm},
          roundedrect/.style={rectangle, rounded corners, draw=black, very thick, minimum size=5mm, align=center},
          node distance=0.2cm and 1.0cm
          ]
          
          \node[roundedrect]              (body1)                         {\texttt{Statements before wait\_for}};
          \node[queue={Context\\FIFO},minimum height=0.5cm]    (context)   [below=2mm of body1]    {};
          \node[roundedrect, text width=10mm]              (expr)      [below=2mm of context]  {\texttt{expr}};
          \node[roundedrect]              (body2)     [below=2mm of expr]     {\texttt{Statements after wait\_for}};
         
          \draw[thick,->] (body1.south) -- (context.north);
          \draw[thick,->] (context.south) -- (expr.north);
          \draw[thick,->] (expr.310) -- ++(0,-0.1) -- ++(1.0,0) |- ++(0,0.77) -| (expr.45);
          \draw[thick,->] (expr.south) -- (body2.north);            
      \end{tikzpicture}    
  \end{subfigure}
  \caption{Hardware realization of \texttt{wait\_for(expr)}.}
  \label{wait_for}
\end{figure*}

\subsubsection{Synchronization - \texttt{wait\_for()}}
As mentioned in Section \ref{schedul_const}, mutex, semaphores, and read-write locks are all \langname{} standard library abstractions built using the \texttt{wait\_for()} primitive.   Figure \ref{wait_for} shows how \texttt{wait\_for()} is implemented in hardware.  Threads execute up to the \texttt{wait\_for(expr)} statement then push the values of thread-local variables into a context FIFO.  The \texttt{wait\_for()} condition can be arbitrarily complex, but is evaluated atomically (\ie{} only for the thread at the head of the FIFO) until it evaluates to true.  When \texttt{expr} is satisfied, the thread continues execution of the statements after the \texttt{wait\_for()}.

\section{Evaluation}
\label{sec:eval}

\subsection{HLS Comparison}
\label{sec:eval-hls}

In this section, we evaluate the \texttt{CountIf Histogram} using two existing HLS tools and the four \langname{} implementations presented in the paper. For these experiments, $N = 512$, $SIZE = 32$, $THRESHOLD = 8$, and $L = 8$. For the HLS implementations, we use Xilinx Vitis HLS 2023.1.1 \cite{xilinx-hls} and Dynamatic v2.0.0 \cite{dynamic-hls-2018}. The Dynamatic and \langname{} implementations are compiled to VHDL and SystemVerilog, respectively, and then synthesized using Xilinx Vivado 2023.1.1. All implementations are synthesized for a 200 MHz clock on the Xilinx Alveo U250 card.

Table \ref{table:hls_vs_ibis} shows the resource usage and latency (best- and worst-case depending on data) for the different implementations.  Vitis HLS implements a static schedule.  Dynamatic \cite{dynamic-hls-2018} implements a dynamic schedule, achieving lower best-case latency but requiring significantly more resources.  This overhead occurs because Dynamatic implements a Load-Store Queue (LSQ).  This structure alone adds 16,803 LUTs and 3,489 FFs. 

The \langname{} Static Scheduling implementation (Figure \ref{StaticScheduleCIH}) has a schedule similar to Vitis HLS, resulting in similar latency. The \langname{} solution uses slightly more resources due to the control FIFOs and memory bypass logic used, but on larger designs these overheads would likely be negligible.  Importantly, \langname{} allows the programmer to express other solutions in the design space.  The Dynamic Scheduling (Figure \ref{DynamicallyScheduledCIH}) and Speculative Execution (Figure \ref{SpeculativeCountIf}) solutions are able to reach lower best-case latency at the cost of more resources and higher worst-case latency. 
In this example, the Dynamic Scheduling implementation is always correct in its decision to obtain the lock, but for other applications where the lock may need to be acquired pessimistically, the Speculative Execution approach can lead to faster execution.  The Dynamic Scheduling solution has worse worst-case latency than Static Scheduling because the most straightforward implementation of the lock adds latency when a conflict occurs (\ie{} the lock is not freed by one thread and acquired by another in the same cycle).  A more sophisticated implementation could reduce this overhead.  Compared to Dynamatic, the \langname{} Dynamic Scheduling implementation has better best-case and worse worst-case latency.  More importantly, though, the resource overhead of Dynamic compared to Static Scheduling is much more reasonable.  The \langname{} Replicated Histogram implementation from Figure \ref{CSlowCIH} uses the most resources among the \langname{} solutions, but has data-independent latency.  The best-case latency is higher compared to the Dynamic Scheduling and Speculative Execution implementations, but the worst-case is significantly lower. Depending on the programmer's requirements, they are able to describe the most suitable implementation.

\subsection{Application Benchmarks}
In this section, we demonstrate that the \langname{} language and toolflow can produce high quality hardware when targeting a variety of ASIC and FPGA platforms.  The baseline for comparison is SystemVerilog implementations for several real-world applications.  This RTL was either in-production or evaluated-for-production, each having multiple person-months to person-years of experienced engineering invested.  These applications were re-implemented using \langname{}.

The \langname{} results are highly competitive for traditional quality metrics such as performance and resource requirements.  Our results are generally comparable to or sometimes better than the hand-written SystemVerilog since many aspects of design-space exploration can be performed without code changes with \langname{} and even architectural changes can often be made easily.  Although the impact of \langname{} on developer effort is difficult to accurately capture quantitatively, we use lines-of-code as a proxy, roughly representing the complexity of initial development, testing, debugging, and future maintenance.  The \langname{} implementations are only 0.2x to 0.5x as many lines of code as the SystemVerilog implementations.

\begin{savenotes}
    \begin{table*}
        \abovecaptionskip = 6pt
        \belowcaptionskip = 10pt
        \caption{Xilinx Vitis HLS, Dynamatic, and \langname{} implementations of \texttt{CountIf Histogram}.}
        \label{table:hls_vs_ibis}
        \scriptsize
        \begin{tabular}{c|c|c|cccc}
        \toprule
                            & Vitis         & Dynamatic     & \multicolumn{4}{c}{\langname{}} \\
                            &               &               &Static Sched.       &Dynamic Sched.      &Replicated Hist.   &Speculative Exec.\\
        \midrule
        LUT                 & 334 (1.00x)   & 17593 (57.67x)& 395 (1.18x)   & 537 (1.61x)   & 825 (2.47x)   & 536 (1.60x)\\
        FF                  & 430 (1.00x)   & 4319 (10.04x) & 608 (1.41x)   & 877 (2.04x)   & 1131 (2.63x)  & 745 (1.73x)\\
        BRAM                & 1 (1.00x)     & 1 (1.00x)     & 1 (1.00x)     & 1 (1.00x)     & 2.5 (2.50x)   & 2.5 (2.50x)\\
        DSP                 & 2 (1.00x)     & 2 (1.00x)     & 2 (1.00x)     & 2 (1.00x)     & 4 (2.00x)     & 2 (1.00x)\\
        \hline
        Best-case Latency          & 4102 (1.00x)         & 1540 (0.38x)          & 4108 (1.00x)         & 530 (0.13x)          & 852 (0.21x)          & 569 (0.14x)\\
        Worst-case Latency         & 4102 (1.00x)         & 4106 (1.00x)         & 4108 (1.00x)         & 5640 (1.37x)         & 852 (0.21x)          & 7723 (1.88x)\\
        \hline
        Lines of Code           & 53 (1.00x)           & 47 (0.89x)           & 60 (1.13x)           & 69 (1.30x)           & 82 (1.55x)           & 123 (2.32x)\\
        \bottomrule
        \end{tabular}
        \caption{ASIC Benchmarks - Post Synthesis Area, Clock Frequency, and Simulation}
        \label{tab:asictable}
            \scriptsize
            \begin{tabular}{c|c|cccc}
              \toprule
              \multicolumn{2}{c|}{}                       &Normalized Area &Target Frequency             &Normalized Throughput   &Lines of Code\\
              \midrule
              \multirow{2}{7.0em}{Decompress}           &RTL        &1.00x      &\SI{1.0}{\giga \hertz}   &1.00x              &8858 (1.00x)\\
                                                        &\langname{}&1.00x      &\SI{1.0}{\giga \hertz}   &1.04x              &4166 (0.49x)\\
              \hline
              \multirow{2}{7.0em}{Match Selection}      &RTL        &1.00x      &\SI{1.2}{\giga \hertz}   &1.00x              &12400 (1.00x)\\
                                                        &\langname{}&1.01x      &\SI{1.2}{\giga \hertz}   &0.60x              &2600 (0.21x)\\
              \bottomrule
          \end{tabular}
          \caption{FPGA Benchmarks - Arria 10 and Stratix 10}
          \label{tab:fpgatable}
          \scriptsize
          \begin{tabular}{c|c|ccccc}
              \toprule
              \multicolumn{2}{c|}{}                                     &ALM                        &BRAM                     &Max Frequency              &Norm. Throughput      &Lines of Code\\
              \midrule
              \multirow{2}{8em}{\centering SDN-A10}     &RTL            &42280 (1.00x)            &1097 (1.00x)               &\SI{200}{\mega \hertz}     &1.00x      &13221 (1.00x)\\
                                                          &\langname{}    &41956 (0.99x)            &992 (0.90x)                &\SI{300}{\mega \hertz}     &4.00x      &5528 (0.42x)\\
              \hline
              \multirow{2}{8em}{\centering Network Prot.-S10}     &RTL            &10133 (1.00x)            &20 (1.00x)                 &\SI{312}{\mega \hertz}     &1.00x      &7228 (1.00x)\\
                                                                    &\langname{}    &9495 (0.94x)             &14 (0.70x)                 &\SI{410}{\mega \hertz}     &1.45x      &2430 (0.33x)\\
              \hline
              \multirow{2}{8em}{\centering AES GCM-S10}           &RTL            &12034 (1.00x)            &161 (1.00x)                &\SI{480}{\mega \hertz}     &1.00x      &4663 (1.00x)\\
                                                                    &\langname{}    &11978 (1.00x)            &163 (1.01x)                &\SI{480}{\mega \hertz}     &1.00x      &1233 (0.26x)\\
          \bottomrule
        \end{tabular}
        \caption{RISC-V Benchmark - Agilex 7}
        \label{tab:riscvtable}
        \scriptsize
        \begin{tabular}{c|ccccc}
            \toprule
            &ALM&BRAM&Max Frequency&CoreMark/MHz&CoreMark\\
            \midrule
            Nios V/m\footnote{Quartus Pro Prime generates encrypted IP for the Nios V/m, the code is not viewable to users.}&1982 (1.00x)&2 (1.00x)&425 MHz&0.40 (1.00x) \cite{intelniosv}&170 (1.00x)\\
            \langname{}&1320 (0.67x)&2 (1.00x)&320 MHz&1.06 (2.65x)&339 (1.99x)\\
        \bottomrule
      \end{tabular}
    \end{table*}
\end{savenotes}

\subsubsection{ASIC Benchmarks}\label{sec:asic-benchmarks}
To show the results for an ASIC platform, we examine two benchmarks targeting two different contemporary technology nodes.  The first benchmark is a DEFLATE decompression module with a 16-bit input and 64-bit output datapath.  This application is control-intensive, primarily because each byte of input can produce a variable number of output bytes.  The second application implements match selection for compression.  A distinctive characteristic of this benchmark is that, as per the baseline architectural design, it is largely single-threaded.  The handwritten SystemVerilog and the SystemVerilog generated by the \langname{} compiler were synthesized using the same production toolflow: a process-specific standard-cell library, Synopsys 2022.03 synthesis tools for regular logic, and a process-specific memory compiler for memory blocks.  Given the nature of ASIC projects, the target clock frequency for both applications were part of the original design specification.  The throughput of all implementations after synthesis was tested using SystemVerilog simulation and internal performance benchmarks.

As shown in Table \ref{tab:asictable}, the \langname{} implementation for both applications met timing and had area nearly identical to the handwritten RTL.  For the decompression benchmark, the throughput of the \langname{} implementation (an average of the performance across an internal suite of tests) was slightly faster than the handwritten RTL, due to minor micro-architectural differences.  Beyond the slight performance advantage, the \langname{} code was also less than half the length of the handwritten SystemVerilog for decompression and only 0.21x for match selection.  

However, the throughput of the match selection benchmark was only 0.6x the handwritten RTL.  
This occurred for two reasons.  First, \langname{} excels at expressing concurrent pipelined computation, but the architectural design and interface of this application was fundamentally single-threaded.
Second, the control structures that are generated by the \langname{} compiler can increase end-to-end latency.
Single-threaded computations are inherently latency sensitive, so these factors combine to decrease performance.
These results are from a preliminary implementation, before planned latency and architectures optimizations were implemented.

\subsubsection{FPGA Benchmarks}\label{sec:fpga-benchmarks}
To show the results when targeting an FPGA, we examined three additional benchmarks implemented on two different Intel FPGAs (Arria 10 and Stratix 10).  The first benchmark is a Software-Defined Networking (SDN) accelerator, responsible for performing 50Gbps packet parsing and modification on an Arria 10.  This application is control-intensive, primarily due to the diversity of packet types it needs to handle.  
The second benchmark is a network protocol engine that generates packet headers and computes CRCs on a Stratix 10.  This application was written in \langname{} using highly parameterized code to enable architectural design-space exploration.  The third benchmark is AES-256 GCM with a 128-bit input and output datapath on a Stratix 10.  This is a dataflow-style application that must run at a comparatively high frequency, which makes effective pipelining important.

As with the handwritten ASIC benchmarks, the baseline for the FPGA benchmarks was in-production or evaluated-for-production SystemVerilog.  Both the handwritten and \langname{} implementations were synthesized using the same toolflow - Quartus Pro 19.1 for the Arria 10 and Quartus Pro Prime 21.4 for the Stratix 10.  However, unlike the ASIC tests we did not fix the same timing constraints for both codebases.  Rather, taking advantage of the much more flexible nature of FPGAs and the ease of generating and using different clocks, we performed a separate timing sweep for each implementation and report the best achievable clock rate for four compilation seeds.

As shown in Table \ref{tab:fpgatable}, the \langname{} implementation of the SDN and protocol engine benchmarks required fewer ALM and block RAM resources while able to run at higher clock rates than the handwritten SystemVerilog.  The source code was also 0.42x and 0.33x the length, respectively.

Perhaps most importantly though, the throughput of these two applications for internal benchmarks was higher.  In the case of the SDN accelerator, this performance advantage was primarily due to two factors.  First, the clock rate was 1.5x higher.  The amount of pipelining in the \langname{} implementation could be changed without source code modifications, facilitating the search for a balance between resource utilization and clock rate.  This increase in clock rate was directly responsible for 1.5x better throughput.  Second, the ease of design-space exploration allowed the \langname{} developers to find a higher throughput micro-architecture, requiring fewer cycles to process tasks.  This was roughly responsible for an additional 2.6x better throughput.  Combined, the \langname{} implementation of the SDN accelerator was 4.0x faster than the handwritten RTL.  Similar clock rate and latency optimizations occurred for the protocol engine benchmark, combining for 1.45x better performance compared to the RTL implementation.

The AES GCM benchmark was nearly identical to the hand-written implementation in terms of both resource requirements and performance, though only requiring 0.26x the lines of code.  That said, the \langname{} implementation did require two additional block memories due to slightly different FIFO arrangements between processing blocks.

We also use \langname{} to implement a RISC-V processor and compare with a slightly different commercially available implementation, both mapped to Intel Agilex 7 FPGA using Quartus Pro Prime 22.4.  The \langname{} RISC-V processor implements a five-stage pipeline for the RV32I ISA.  The Intel Nios V/m also implements a five-stage pipeline, but for the RV32IZicsr ISA, which adds control and status register instructions. Looking at the performance and resource requirements in Table \ref{tab:riscvtable}, it is obvious that the processors have different architectural goals.  Although this comparison is imperfect, it shows that \langname{} can be used to implement a performant RISC-V processor with reasonable resource requirements.  This is significant because this is generally not true for traditional HLS tools\cite{hl5-2020}.  For example, without constructs to explicitly define concurrency, it can be difficult to express features such as control and data hazard handling, which are specific to how the processor is divided into pipeline stages.

\subsubsection{Academic Research}\label{sec:academic}
\langname{} has also been used in two recent papers.  The first was \citet{heax-2020}, which used \langname{} to build an FPGA implementation of a Fully Homomorphic Encryption algorithm.  Beyond implementing the cryptographic system itself (\eg{} polynomial modulo math, tuning the design to the available DSP and memory blocks), system-level optimizations were added (\eg{} batching data and multi-buffering communication with the host machine to hide PCIe latency).  The \langname{} FPGA implementation was faster than the Microsoft software library implementation.  The second project to use \langname{} was \citet{pqc-2023}, which accelerated a post-quantum key encapsulation mechanism.  The authors built heavily parameterized \langname{} code, allowing them to easily generate customized platforms.  The \langname{} FPGA implementation was compelling compared to previous work and required less than 0.20x the development time of a manual RTL implementation.

\section{Related Work}
\label{sec:related}
In this section, we qualitatively compare \langname{} with other high-level hardware languages. 

\subsection{High-level HDL}
Bluespec SystemVerilog \cite{bluespec-2004} and Kôika \cite{koika-2020} offer more comprehensible semantics than RTL languages by modeling hardware execution as the application of a set of atomic rules.  Each computational step logically selects exactly one rule and applies it, giving one-rule-at-a-time (ORAAT) semantics.  To improve hardware performance, dynamic scheduling is used to enable multiple rules to run in parallel, when parallel execution is equivalent to ORAAT execution.

Languages like Chisel \cite{chisel-dac12} embed hardware description within a software language.  Constructs from the host language can be used to algorithmically generate hardware descriptions.

While these languages raise the abstraction level compared to RTL, they still retain the low-level hardware programming paradigm.  State elements are explicitly specified in these languages, along with transition functions or rules that complete on short time scales.  Source code in these languages is less abstract than typical \langname{} source and does not use imperative control flow.  

\subsection{Traditional HLS}
As discussed in Section \ref{sec:eval-hls}, high-level synthesis tools such as Vitis HLS \cite{xilinx-hls}, Intel HLS Compiler \cite{intel-hls}, LegUp \cite{legup-2013}, Catapult-C \cite{catapult-c}, and Cadence Stratus \cite{stratus} take on the difficult task of transforming sequential source into parallel hardware.

\subsubsection{Message Passing}
An execution model supported by HLS tools is explicit task parallelism described as Kahn Process Networks \cite{kahn-1974} or Communicating Sequential Processes \cite{Hoare-1978}.  Tasks communicate through message passing.  Each state element (except for message queues) is assigned to a specific task which is the only task that can directly access that state element.  Concurrency between separate tasks is straightforward to map to coarse-grain hardware parallelism, while automatically extracting fine-grained parallelism from within a single sequential task remains difficult.  This leads seasoned HLS engineers to express hardware with a fine-grained decomposition into many (small) task-parallel modules.

As the decomposition becomes more fine-grained, it becomes difficult to assign data structures to tasks.  For example, each pipeline stage in a processor can be described as a separate task, sending and receiving messages to and from other stages.  However, state elements such as the register file and program counter must be assigned to a single task which can access them directly.  This can lead to sub-optimal implementations where instruction fetch, decode, and register reads all occur in a single task \cite{hl5-2020}.  Another example is a network packet processor modeled as a set of parallel tasks which send packets as messages to one another.  It is impossible to atomically update configuration data structures distributed throughout the design without draining all in-flight packets \cite{xilinx-abstract-parallel}. 

Figure \ref{ResolveSort} shows idiomatic examples of message passing HLS from the literature \cite{resolve-2016}.  \citet{resolve-2016} begins with a discussion of how HLS tools cannot efficiently implement sorting algorithms described imperatively, and demonstrates how message passing produces better results.  Distinguishing features of message passing HLS include:

\begin{itemize}
    \item Function parameters are always queues (\texttt{hls::stream} in this case).
    \item Function calls occur unconditionally.
    \item Imperative control flow is used sparingly, with control implemented as data flow.
\end{itemize}

\langname{} thread synchronization and communication mechanisms allow the expression of a wider range of designs than message passing HLS because \langname{} threads can access shared state directly. 

\begin{figure}[t]
  \belowcaptionskip = 0pt
  \begin{minipage}[t]{.75\linewidth}
    \begin{lstlisting}[language=HLS]
T InsertionCell(hls::stream <int> &IN, hls::stream <int> &OUT){
  static int CURR_REG=0;
  int IN_A = IN.read();
  if(IN_A > CURR_REG) {
      OUT.write(CURR_REG); 
      CURR_REG = IN_A;
  } else
      OUT.write(IN_A);
  return CURR_REG;
}
void InsertionSort(hls::stream <T> &IN, hls::stream <T> &OUT){
#pragma HLS DATAFLOW
  hls::stream <T> out1 , out2 , out3;
  // Function calls;
  InsertionCell1(IN, out1);
  InsertionCell2(out1, out2);
  InsertionCell3(out2, out3);
  InsertionCell4(out3, OUT);
}
    \end{lstlisting}
  \caption{Sorting modules implemented with message passing.}
  \label{ResolveSort}
  \end{minipage}
\end{figure}

\subsubsection{Shared Memory Concurrency}
There has been some work on shared memory concurrency for HLS.  LegUp \cite{legup-threads-2013} supports expressing coarse-grain concurrency with threads that behave like typical software threads.  Threads are mapped to parallel hardware modules which can communicate through shared state.  \citet{weak-consistent-hls-2017} and \citet{concurrency-aware-2018} describe modifications to LegUp to support the semantics of C atomics.   OpenCL supports shared memory concurrency through lightweight threads executing computational kernels, and synchronizing with constructs which map well to GPUs \cite{opencl-2012}.

Rather than inheriting shared memory concurrency semantics from languages that map well to CPUs or GPUs, \langname{} introduces semantics tailored for the reliable generation of efficient hardware.

\subsubsection{Dynamic HLS}
\label{sec:dynamic_hls}
Dynamic HLS \cite{dynamic-hls-2018} implements the same sequential semantics as traditional HLS, but moves much of the responsibility of scheduling from the compiler to the hardware.  Handshaking connections and load-store queues \cite{lsq-2017} in the generated hardware ensure that operations run in an order that maintains sequential semantics.  Dynamic scheduling outperforms static scheduling in cases where a static scheduler has to make overly conservative decisions to accommodate worst-case behavior.  As with traditional HLS, the constraint of implementing a sequential source language can lead to situations where the compiler must perform non-trivial transformations \cite{c-slow-dynamic-2022, inter-block-2022}  in order to achieve high throughput.

As we have shown with the \texttt{CountIf Histogram} example, \langname{} allows the programmer to explicitly express a variety of solutions, each having different design trade-offs.

\subsection{Spatial}
Spatial \cite{spatial-2018} is an imperative, domain-specific language for application accelerators.  Control is expressed with the composition of a set of control structures including parallel patterns like \texttt{Foreach} and \texttt{Reduce}.  Like traditional HLS compilers, the Spatial compiler only pipelines loops when the compiler can ensure that the behavior of a pipelined loop matches the behavior of sequential implementation. \langname{} allows the programmer to pipeline arbitrary code and gives the programmer scheduling constraints and synchronization tools to ensure correctness.

\subsection{PDL}
PDL \cite{pdl-pldi2022} models a processor as a stream of threads flowing down a manually-scheduled pipeline, with one thread handling the execution of a single instruction.  Hazards are resolved with the help of synchronization objects which are referenced by PDL code and implemented in RTL.  \langname{} generalizes this notion of multi-threading in the following ways:

\begin{itemize}
\item \langname{} designs compile to hardware comprising multiple pipelines, with imperative control flow dictating how threads move between pipelines.
\item \langname{} designs are automatically pipelined, subject to user-specified constraints.
\item \langname{} is sufficiently general that synchronization objects can be implemented in \langname{} directly.
\end{itemize}

\subsection{Dahlia}
Dahlia \cite{dahlia-2020} is a novel HLS language in which the central tenet is that it only attempts to compile programs with predictable performance to hardware.  It accomplishes this by defining a type system that, when met, allows the compiler to directly map computations to hardware.  Programs that include potentially complex or unpredictable constructs, such as arbitrary array indexing, fail type checking.  \langname{} also aims for predictable performance, but achieves this by requiring the programmer to explicitly describe concurrency.

\subsection{Calyx}
Calyx \cite{calyx-2021} is an intermediate language which combines structural and imperative composition. Data paths and connectivity are described structurally, while the control logic for each module is defined imperatively. Calyx and \langname{} model imperative concurrency differently. In Calyx, disjoint subsets of code (“groups”) can execute in parallel, whereas in \langname{} multiple threads can be executing the same code concurrently (each with separate thread-local variables).

\subsection{Filament}
Filament \cite{filament-2023} is a hardware description language with timeline types. The Filament type checker can catch bugs related to the composition of hardware modules such as the possibility of reading the output of a module at the incorrect clock cycle. Like other higher level HDLs, Filament raises the abstraction level compared to RTL, but is still low level relative to \langname{}.

\section{Limitations of \langname{}}
\label{sec:limitations}
In our experience, \langname{} is an easy-to-use, highly capable language with distinct advantages over existing alternatives.  
For example, several interns with only prior software experience have been able to learn \langname{} and become proficient enough to create efficient FPGA implementations of non-trivial applications in fewer than 12 weeks.
That said, hardware development is a long-standing and multi-faceted problem.  We cannot claim that \langname{} is the most appropriate tool for all situations.  Compared to traditional HLS, \langname{} relies on the programmer to discover and express concurrency with threads.  Single-threaded code results in poor throughput and multi-threaded code often requires explicit synchronization to ensure correct behavior. Although \consistname{} makes reasoning about concurrency easier than in typical multi-threaded programs, subtle synchronization errors can result in elusive bugs.  We believe that formal verification could help to catch these bugs.  

\langname{} is also not as expressive as RTL. For example, tight coordination between threads is only possible for ordered threads that are executing the same statements.  Threads executing disjoint regions of code cannot be as tightly coordinated.
Also, as shown in the Match Selection benchmark in \ref{sec:asic-benchmarks}, it can be difficult to precisely control the latency of the generated output.  This can affect single-threaded performance.

\section{Conclusion}
\label{sec:conclusion}

\langname{} empowers programmers to build hardware designs with a highly expressive language, leveraging familiar programming language features that make code naturally composable and reusable. \langname{} supports an execution model and a memory consistency model that make generating efficient hardware simple for the compiler and reduces the complexity of reasoning about highly concurrent code. In our experience, the language is easy to learn and new developers can become productive within weeks.  \langname{} also permits developers to iterate on complex designs with low friction. They can rapidly experiment with architectural choices to optimize performance and resource requirements.  We have shown that \langname{} can produce high-quality circuits in a production environment.

Looking ahead, generative AI for software development is a nascent area showing immense promise.  We believe that \langname{} is well poised as a platform for generative hardware development.  The same characteristics that make it excel for manual development (\eg{} compact syntax, zero-cost abstractions, functional composability, and a strong, static type system) make it a highly practical target for large language AI models.  Both human and AI \langname{} developers can focus on the higher-level functional semantics because the \langname{} compiler takes care of lower-level implementation details.  This encourages more advanced, efficient, powerful, and reliable hardware solutions.

\begin{acks}
    The authors thank the anonymous reviewers for their valuable feedback, suggestions, and insights that have significantly improved the quality of this paper. We also express our gratitude to esteemed academic partners for their collaboration and intellectual contributions, which have been crucial in developing this novel hardware language.  The authors thank Mark Hill for his in-depth discussions and suggestion to frame the description of the \langname{} semantics as a memory consistency model.
\end{acks}

\bibliographystyle{ACM-Reference-Format}

\bibliography{paper}


\begin{thebibliography}{56}


\ifx \showCODEN    \undefined \def \showCODEN     #1{\unskip}     \fi
\ifx \showDOI      \undefined \def \showDOI       #1{#1}\fi
\ifx \showISBNx    \undefined \def \showISBNx     #1{\unskip}     \fi
\ifx \showISBNxiii \undefined \def \showISBNxiii  #1{\unskip}     \fi
\ifx \showISSN     \undefined \def \showISSN      #1{\unskip}     \fi
\ifx \showLCCN     \undefined \def \showLCCN      #1{\unskip}     \fi
\ifx \shownote     \undefined \def \shownote      #1{#1}          \fi
\ifx \showarticletitle \undefined \def \showarticletitle #1{#1}   \fi
\ifx \showURL      \undefined \def \showURL       {\relax}        \fi
\providecommand\bibfield[2]{#2}
\providecommand\bibinfo[2]{#2}
\providecommand\natexlab[1]{#1}
\providecommand\showeprint[2][]{arXiv:#2}

\bibitem[CIR(2023)]%
        {CIRCT}
 \bibinfo{year}{2023}\natexlab{}.
\newblock \bibinfo{title}{CIRCT}.
\newblock \bibinfo{howpublished}{\url{https://circt.llvm.org/}}.
\newblock
\newblock
\shownote{(accessed: 11.8.2023)}.


\bibitem[Allen and Cocke(1976)]%
        {dfa-1973}
\bibfield{author}{\bibinfo{person}{F.~E. Allen} {and} \bibinfo{person}{J.
  Cocke}.} \bibinfo{year}{1976}\natexlab{}.
\newblock \showarticletitle{A Program Data Flow Analysis Procedure}.
\newblock \bibinfo{journal}{\emph{Commun. ACM}} \bibinfo{volume}{19},
  \bibinfo{number}{3} (\bibinfo{date}{mar} \bibinfo{year}{1976}),
  \bibinfo{pages}{137}.
\newblock
\showISSN{0001-0782}
\urldef\tempurl%
\url{https://doi.org/10.1145/360018.360025}
\showDOI{\tempurl}


\bibitem[Bachrach et~al\mbox{.}(2012)]%
        {chisel-dac12}
\bibfield{author}{\bibinfo{person}{Jonathan Bachrach}, \bibinfo{person}{Huy
  Vo}, \bibinfo{person}{Brian Richards}, \bibinfo{person}{Yunsup Lee},
  \bibinfo{person}{Andrew Waterman}, \bibinfo{person}{Rimas Avižienis},
  \bibinfo{person}{John Wawrzynek}, {and} \bibinfo{person}{Krste Asanović}.}
  \bibinfo{year}{2012}\natexlab{}.
\newblock \showarticletitle{Chisel: Constructing hardware in a Scala embedded
  language}. In \bibinfo{booktitle}{\emph{DAC Design Automation Conference
  2012}}. \bibinfo{pages}{1212--1221}.
\newblock
\urldef\tempurl%
\url{https://doi.org/10.1145/2228360.2228584}
\showDOI{\tempurl}


\bibitem[Bisheh-Niasar et~al\mbox{.}(2023)]%
        {pqc-2023}
\bibfield{author}{\bibinfo{person}{Mojtaba Bisheh-Niasar},
  \bibinfo{person}{Daniel Lo}, \bibinfo{person}{Anjana Parthasarathy},
  \bibinfo{person}{Blake Pelton}, \bibinfo{person}{Bharat Pillilli}, {and}
  \bibinfo{person}{Bryan Kelly}.} \bibinfo{year}{2023}\natexlab{}.
\newblock \showarticletitle{PQC Cloudization: Rapid Prototyping of Scalable NTT
  /INTT Architecture to Accelerate Kyber}. In \bibinfo{booktitle}{\emph{2023
  IEEE Physical Assurance and Inspection of Electronics (PAINE)}}.
  \bibinfo{pages}{1--7}.
\newblock
\urldef\tempurl%
\url{https://doi.org/10.1109/PAINE58317.2023.10318029}
\showDOI{\tempurl}


\bibitem[Bourgeat et~al\mbox{.}(2020)]%
        {koika-2020}
\bibfield{author}{\bibinfo{person}{Thomas Bourgeat},
  \bibinfo{person}{Cl\'{e}ment Pit-Claudel}, \bibinfo{person}{Adam Chlipala},
  {and} \bibinfo{person}{Arvind}.} \bibinfo{year}{2020}\natexlab{}.
\newblock \showarticletitle{The Essence of Bluespec: A Core Language for
  Rule-Based Hardware Design}. In \bibinfo{booktitle}{\emph{Proceedings of the
  41st ACM SIGPLAN Conference on Programming Language Design and
  Implementation}} (London, UK) \emph{(\bibinfo{series}{PLDI 2020})}.
  \bibinfo{publisher}{Association for Computing Machinery},
  \bibinfo{address}{New York, NY, USA}, \bibinfo{pages}{243–257}.
\newblock
\showISBNx{9781450376136}
\urldef\tempurl%
\url{https://doi.org/10.1145/3385412.3385965}
\showDOI{\tempurl}


\bibitem[{Cadence Design Systems Inc.}(2023)]%
        {stratus}
\bibfield{author}{\bibinfo{person}{{Cadence Design Systems Inc.}}}
  \bibinfo{year}{2023}\natexlab{}.
\newblock \bibinfo{title}{Stratus High-Level Synthesis}.
\newblock
  \bibinfo{howpublished}{\url{https://www.cadence.com/en_US/home/tools/digital-design-and-signoff/synthesis/stratus-high-level-synthesis.html}}.
\newblock
\newblock
\shownote{(accessed: 11.2.2023)}.


\bibitem[Canis et~al\mbox{.}(2013)]%
        {legup-2013}
\bibfield{author}{\bibinfo{person}{Andrew Canis}, \bibinfo{person}{Jongsok
  Choi}, \bibinfo{person}{Mark Aldham}, \bibinfo{person}{Victor Zhang},
  \bibinfo{person}{Ahmed Kammoona}, \bibinfo{person}{Tomasz Czajkowski},
  \bibinfo{person}{Stephen~D. Brown}, {and} \bibinfo{person}{Jason~H.
  Anderson}.} \bibinfo{year}{2013}\natexlab{}.
\newblock \showarticletitle{LegUp: An Open-Source High-Level Synthesis Tool for
  FPGA-Based Processor/Accelerator Systems}.
\newblock \bibinfo{journal}{\emph{ACM Trans. Embed. Comput. Syst.}}
  \bibinfo{volume}{13}, \bibinfo{number}{2}, Article \bibinfo{articleno}{24}
  (\bibinfo{date}{sep} \bibinfo{year}{2013}), \bibinfo{numpages}{27}~pages.
\newblock
\showISSN{1539-9087}
\urldef\tempurl%
\url{https://doi.org/10.1145/2514740}
\showDOI{\tempurl}


\bibitem[Chamberlain et~al\mbox{.}(1999)]%
        {chamberlain1999array}
\bibfield{author}{\bibinfo{person}{Bradford~L Chamberlain},
  \bibinfo{person}{E~Christopher Lewis}, {and} \bibinfo{person}{Lawrence
  Snyder}.} \bibinfo{year}{1999}\natexlab{}.
\newblock \showarticletitle{Array language support for wavefront and pipelined
  computations}. In \bibinfo{booktitle}{\emph{Workshop on Languages and
  Compilers for Parallel Computing}}. Citeseer.
\newblock


\bibitem[Cheng et~al\mbox{.}(2022a)]%
        {inter-block-2022}
\bibfield{author}{\bibinfo{person}{Jianyi Cheng}, \bibinfo{person}{Lana
  Josipović}, \bibinfo{person}{George~A. Constantinides}, {and}
  \bibinfo{person}{John Wickerson}.} \bibinfo{year}{2022}\natexlab{a}.
\newblock \showarticletitle{Dynamic Inter-Block Scheduling for HLS}. In
  \bibinfo{booktitle}{\emph{2022 32nd International Conference on
  Field-Programmable Logic and Applications (FPL)}}. \bibinfo{pages}{243--252}.
\newblock
\urldef\tempurl%
\url{https://doi.org/10.1109/FPL57034.2022.00045}
\showDOI{\tempurl}


\bibitem[Cheng et~al\mbox{.}(2021)]%
        {cheng-2021}
\bibfield{author}{\bibinfo{person}{Jianyi Cheng}, \bibinfo{person}{John
  Wickerson}, {and} \bibinfo{person}{George~A. Constantinides}.}
  \bibinfo{year}{2021}\natexlab{}.
\newblock \showarticletitle{Exploiting the Correlation between Dependence
  Distance and Latency in Loop Pipelining for HLS}. In
  \bibinfo{booktitle}{\emph{2021 31st International Conference on
  Field-Programmable Logic and Applications (FPL)}}. \bibinfo{pages}{341--346}.
\newblock
\urldef\tempurl%
\url{https://doi.org/10.1109/FPL53798.2021.00066}
\showDOI{\tempurl}


\bibitem[Cheng et~al\mbox{.}(2022b)]%
        {c-slow-dynamic-2022}
\bibfield{author}{\bibinfo{person}{Jianyi Cheng}, \bibinfo{person}{John
  Wickerson}, {and} \bibinfo{person}{George~A. Constantinides}.}
  \bibinfo{year}{2022}\natexlab{b}.
\newblock \showarticletitle{Dynamic C-Slow Pipelining for HLS}. In
  \bibinfo{booktitle}{\emph{2022 IEEE 30th Annual International Symposium on
  Field-Programmable Custom Computing Machines (FCCM)}}.
  \bibinfo{pages}{1--10}.
\newblock
\urldef\tempurl%
\url{https://doi.org/10.1109/FCCM53951.2022.9786096}
\showDOI{\tempurl}


\bibitem[Choi et~al\mbox{.}(2013)]%
        {legup-threads-2013}
\bibfield{author}{\bibinfo{person}{Jongsok Choi}, \bibinfo{person}{Stephen
  Brown}, {and} \bibinfo{person}{Jason Anderson}.}
  \bibinfo{year}{2013}\natexlab{}.
\newblock \showarticletitle{From software threads to parallel hardware in
  high-level synthesis for FPGAs}. In \bibinfo{booktitle}{\emph{2013
  International Conference on Field-Programmable Technology (FPT)}}.
  \bibinfo{pages}{270--277}.
\newblock
\urldef\tempurl%
\url{https://doi.org/10.1109/FPT.2013.6718365}
\showDOI{\tempurl}


\bibitem[Cong and Zhang(2006)]%
        {sdc-2006}
\bibfield{author}{\bibinfo{person}{J. Cong} {and} \bibinfo{person}{Zhiru
  Zhang}.} \bibinfo{year}{2006}\natexlab{}.
\newblock \showarticletitle{An efficient and versatile scheduling algorithm
  based on SDC formulation}. In \bibinfo{booktitle}{\emph{2006 43rd ACM/IEEE
  Design Automation Conference}}. \bibinfo{pages}{433--438}.
\newblock
\urldef\tempurl%
\url{https://doi.org/10.1145/1146909.1147025}
\showDOI{\tempurl}


\bibitem[Corporation(2023)]%
        {intelniosv}
\bibfield{author}{\bibinfo{person}{Intel Corporation}.}
  \bibinfo{year}{2023}\natexlab{}.
\newblock \bibinfo{booktitle}{\emph{Nios® V Processor Reference Manual -
  Updated for Intel® Quartus® Prime Design Suite: 23.3}}.
\newblock


\bibitem[Czajkowski et~al\mbox{.}(2012)]%
        {opencl-2012}
\bibfield{author}{\bibinfo{person}{Tomasz~S Czajkowski}, \bibinfo{person}{David
  Neto}, \bibinfo{person}{Michael Kinsner}, \bibinfo{person}{Utku Aydonat},
  \bibinfo{person}{Jason Wong}, \bibinfo{person}{Dmitry Denisenko},
  \bibinfo{person}{Peter Yiannacouras}, \bibinfo{person}{John Freeman},
  \bibinfo{person}{Deshanand~P Singh}, {and} \bibinfo{person}{Stephen~D
  Brown}.} \bibinfo{year}{2012}\natexlab{}.
\newblock \showarticletitle{OpenCL for FPGAs: Prototyping a compiler}. In
  \bibinfo{booktitle}{\emph{Proceedings of the International Conference on
  Engineering of Reconfigurable Systems and Algorithms (ERSA)}}. The World
  Congress in Computer Science, Computer Engineering, \& Applied Computing.
\newblock


\bibitem[Dai et~al\mbox{.}(2017)]%
        {dai-fpga17}
\bibfield{author}{\bibinfo{person}{Steve Dai}, \bibinfo{person}{Ritchie Zhao},
  \bibinfo{person}{Gai Liu}, \bibinfo{person}{Shreesha Srinath},
  \bibinfo{person}{Udit Gupta}, \bibinfo{person}{Christopher Batten}, {and}
  \bibinfo{person}{Zhiru Zhang}.} \bibinfo{year}{2017}\natexlab{}.
\newblock \showarticletitle{Dynamic Hazard Resolution for Pipelining Irregular
  Loops in High-Level Synthesis}. In \bibinfo{booktitle}{\emph{Proceedings of
  the 2017 ACM/SIGDA International Symposium on Field-Programmable Gate
  Arrays}} (Monterey, California, USA) \emph{(\bibinfo{series}{FPGA '17})}.
  \bibinfo{publisher}{Association for Computing Machinery},
  \bibinfo{address}{New York, NY, USA}, \bibinfo{pages}{189–194}.
\newblock
\showISBNx{9781450343541}
\urldef\tempurl%
\url{https://doi.org/10.1145/3020078.3021754}
\showDOI{\tempurl}


\bibitem[Derrien et~al\mbox{.}(2020)]%
        {spec-2020}
\bibfield{author}{\bibinfo{person}{Steven Derrien}, \bibinfo{person}{Thibaut
  Marty}, \bibinfo{person}{Simon Rokicki}, {and} \bibinfo{person}{Tomofumi
  Yuki}.} \bibinfo{year}{2020}\natexlab{}.
\newblock \showarticletitle{Toward Speculative Loop Pipelining for High-Level
  Synthesis}.
\newblock \bibinfo{journal}{\emph{IEEE Transactions on Computer-Aided Design of
  Integrated Circuits and Systems}} \bibinfo{volume}{39}, \bibinfo{number}{11}
  (\bibinfo{year}{2020}), \bibinfo{pages}{4229--4239}.
\newblock
\urldef\tempurl%
\url{https://doi.org/10.1109/TCAD.2020.3012866}
\showDOI{\tempurl}


\bibitem[Ferrante et~al\mbox{.}(1987)]%
        {Ferrante-1987}
\bibfield{author}{\bibinfo{person}{Jeanne Ferrante}, \bibinfo{person}{Karl~J.
  Ottenstein}, {and} \bibinfo{person}{Joe~D. Warren}.}
  \bibinfo{year}{1987}\natexlab{}.
\newblock \showarticletitle{The Program Dependence Graph and Its Use in
  Optimization}.
\newblock \bibinfo{journal}{\emph{ACM Trans. Program. Lang. Syst.}}
  \bibinfo{volume}{9}, \bibinfo{number}{3} (\bibinfo{date}{jul}
  \bibinfo{year}{1987}), \bibinfo{pages}{319–349}.
\newblock
\showISSN{0164-0925}
\urldef\tempurl%
\url{https://doi.org/10.1145/24039.24041}
\showDOI{\tempurl}


\bibitem[Gorius et~al\mbox{.}(2022)]%
        {spec-2022}
\bibfield{author}{\bibinfo{person}{Jean-Michel Gorius}, \bibinfo{person}{Simon
  Rokicki}, {and} \bibinfo{person}{Steven Derrien}.}
  \bibinfo{year}{2022}\natexlab{}.
\newblock \showarticletitle{SpecHLS: Speculative Accelerator Design Using
  High-Level Synthesis}.
\newblock \bibinfo{journal}{\emph{IEEE Micro}} \bibinfo{volume}{42},
  \bibinfo{number}{5} (\bibinfo{year}{2022}), \bibinfo{pages}{99--107}.
\newblock
\urldef\tempurl%
\url{https://doi.org/10.1109/MM.2022.3188136}
\showDOI{\tempurl}


\bibitem[Harold~Abelson(1985)]%
        {cisp-1985}
\bibfield{author}{\bibinfo{person}{Gerald Jay~Sussman Harold~Abelson}.}
  \bibinfo{year}{1985}\natexlab{}.
\newblock \bibinfo{booktitle}{\emph{Structure and Interpretation of Computer
  Programs}}.
\newblock \bibinfo{publisher}{The MIT Press}.
\newblock
\showISBNx{9780262510363}


\bibitem[Hoare(1978)]%
        {Hoare-1978}
\bibfield{author}{\bibinfo{person}{C.~A.~R. Hoare}.}
  \bibinfo{year}{1978}\natexlab{}.
\newblock \showarticletitle{Communicating Sequential Processes}.
\newblock \bibinfo{journal}{\emph{Commun. ACM}} \bibinfo{volume}{21},
  \bibinfo{number}{8} (\bibinfo{date}{aug} \bibinfo{year}{1978}),
  \bibinfo{pages}{666–677}.
\newblock
\showISSN{0001-0782}
\urldef\tempurl%
\url{https://doi.org/10.1145/359576.359585}
\showDOI{\tempurl}


\bibitem[{Intel Corporation}(2023)]%
        {intel-hls}
\bibfield{author}{\bibinfo{person}{{Intel Corporation}}.}
  \bibinfo{year}{2023}\natexlab{}.
\newblock \bibinfo{title}{Intel High Level Synthesis Compiler}.
\newblock
  \bibinfo{howpublished}{\url{https://www.intel.com/content/www/us/en/software/programmable/quartus-prime/hls-compiler.html}}.
\newblock
\newblock
\shownote{(accessed: 11.1.2023)}.


\bibitem[Josipovic et~al\mbox{.}(2017)]%
        {lsq-2017}
\bibfield{author}{\bibinfo{person}{Lana Josipovic}, \bibinfo{person}{Philip
  Brisk}, {and} \bibinfo{person}{Paolo Ienne}.}
  \bibinfo{year}{2017}\natexlab{}.
\newblock \showarticletitle{An Out-of-Order Load-Store Queue for Spatial
  Computing}.
\newblock \bibinfo{journal}{\emph{ACM Trans. Embed. Comput. Syst.}}
  \bibinfo{volume}{16}, \bibinfo{number}{5s}, Article \bibinfo{articleno}{125}
  (\bibinfo{date}{sep} \bibinfo{year}{2017}), \bibinfo{numpages}{19}~pages.
\newblock
\showISSN{1539-9087}
\urldef\tempurl%
\url{https://doi.org/10.1145/3126525}
\showDOI{\tempurl}


\bibitem[Josipovi\'{c} et~al\mbox{.}(2018)]%
        {dynamic-hls-2018}
\bibfield{author}{\bibinfo{person}{Lana Josipovi\'{c}},
  \bibinfo{person}{Radhika Ghosal}, {and} \bibinfo{person}{Paolo Ienne}.}
  \bibinfo{year}{2018}\natexlab{}.
\newblock \showarticletitle{Dynamically Scheduled High-Level Synthesis}. In
  \bibinfo{booktitle}{\emph{Proceedings of the 2018 ACM/SIGDA International
  Symposium on Field-Programmable Gate Arrays}} (Monterey, CALIFORNIA, USA)
  \emph{(\bibinfo{series}{FPGA '18})}. \bibinfo{publisher}{Association for
  Computing Machinery}, \bibinfo{address}{New York, NY, USA},
  \bibinfo{pages}{127–136}.
\newblock
\showISBNx{9781450356145}
\urldef\tempurl%
\url{https://doi.org/10.1145/3174243.3174264}
\showDOI{\tempurl}


\bibitem[Josipovic et~al\mbox{.}(2019)]%
        {spec-19}
\bibfield{author}{\bibinfo{person}{Lana Josipovic}, \bibinfo{person}{Andrea
  Guerrieri}, {and} \bibinfo{person}{Paolo Ienne}.}
  \bibinfo{year}{2019}\natexlab{}.
\newblock \showarticletitle{Speculative Dataflow Circuits}. In
  \bibinfo{booktitle}{\emph{Proceedings of the 2019 ACM/SIGDA International
  Symposium on Field-Programmable Gate Arrays}} (Seaside, CA, USA)
  \emph{(\bibinfo{series}{FPGA '19})}. \bibinfo{publisher}{Association for
  Computing Machinery}, \bibinfo{address}{New York, NY, USA},
  \bibinfo{pages}{162–171}.
\newblock
\showISBNx{9781450361378}
\urldef\tempurl%
\url{https://doi.org/10.1145/3289602.3293914}
\showDOI{\tempurl}


\bibitem[Jr.(1976)]%
        {lambda-declarative-1976}
\bibfield{author}{\bibinfo{person}{Guy Lewis~Steele Jr.}}
  \bibinfo{year}{1976}\natexlab{}.
\newblock \showarticletitle{LAMBDA: The Ultimate Declarative}.
\newblock  (\bibinfo{year}{1976}), \bibinfo{pages}{48}.
\newblock
\urldef\tempurl%
\url{http://hdl.handle.net/1721.1/6091}
\showURL{%
\tempurl}


\bibitem[Kahn(1974)]%
        {kahn-1974}
\bibfield{author}{\bibinfo{person}{Gilles Kahn}.}
  \bibinfo{year}{1974}\natexlab{}.
\newblock \bibinfo{title}{The semantics of a simple language for parallel
  programming, IFIP 74}.
\newblock
\newblock


\bibitem[Koeplinger et~al\mbox{.}(2018)]%
        {spatial-2018}
\bibfield{author}{\bibinfo{person}{David Koeplinger}, \bibinfo{person}{Matthew
  Feldman}, \bibinfo{person}{Raghu Prabhakar}, \bibinfo{person}{Yaqi Zhang},
  \bibinfo{person}{Stefan Hadjis}, \bibinfo{person}{Ruben Fiszel},
  \bibinfo{person}{Tian Zhao}, \bibinfo{person}{Luigi Nardi},
  \bibinfo{person}{Ardavan Pedram}, \bibinfo{person}{Christos Kozyrakis}, {and}
  \bibinfo{person}{Kunle Olukotun}.} \bibinfo{year}{2018}\natexlab{}.
\newblock \showarticletitle{Spatial: A Language and Compiler for Application
  Accelerators}. In \bibinfo{booktitle}{\emph{Proceedings of the 39th ACM
  SIGPLAN Conference on Programming Language Design and Implementation}}
  (Philadelphia, PA, USA) \emph{(\bibinfo{series}{PLDI 2018})}.
  \bibinfo{publisher}{Association for Computing Machinery},
  \bibinfo{address}{New York, NY, USA}, \bibinfo{pages}{296–311}.
\newblock
\showISBNx{9781450356985}
\urldef\tempurl%
\url{https://doi.org/10.1145/3192366.3192379}
\showDOI{\tempurl}


\bibitem[Kung et~al\mbox{.}(1982)]%
        {wavefront-1982}
\bibfield{author}{\bibinfo{person}{Sun-Yuan Kung}, \bibinfo{person}{Arun},
  \bibinfo{person}{Gal-Ezer}, {and} \bibinfo{person}{Bhaskar Rao}.}
  \bibinfo{year}{1982}\natexlab{}.
\newblock \showarticletitle{Wavefront Array Processor: Language, Architecture,
  and Applications}.
\newblock \bibinfo{journal}{\emph{IEEE Trans. Comput.}} \bibinfo{volume}{C-31},
  \bibinfo{number}{11} (\bibinfo{year}{1982}), \bibinfo{pages}{1054--1066}.
\newblock
\urldef\tempurl%
\url{https://doi.org/10.1109/TC.1982.1675922}
\showDOI{\tempurl}


\bibitem[Lamport(1997)]%
        {lamport-1997}
\bibfield{author}{\bibinfo{person}{L. Lamport}.}
  \bibinfo{year}{1997}\natexlab{}.
\newblock \showarticletitle{How to make a correct multiprocess program execute
  correctly on a multiprocessor}.
\newblock \bibinfo{journal}{\emph{IEEE Trans. Comput.}} \bibinfo{volume}{46},
  \bibinfo{number}{7} (\bibinfo{year}{1997}), \bibinfo{pages}{779--782}.
\newblock
\urldef\tempurl%
\url{https://doi.org/10.1109/12.599898}
\showDOI{\tempurl}


\bibitem[Ledgard(1983)]%
        {ada-1983}
\bibfield{author}{\bibinfo{person}{Henry Ledgard}.}
  \bibinfo{year}{1983}\natexlab{}.
\newblock \bibinfo{booktitle}{\emph{Reference Manual for the ADA Programming
  Language}}.
\newblock \bibinfo{publisher}{Springer-Verlag}, \bibinfo{address}{Berlin,
  Heidelberg}.
\newblock
\showISBNx{0387908870}


\bibitem[Liu et~al\mbox{.}(2017)]%
        {liu-cadics17}
\bibfield{author}{\bibinfo{person}{Gai Liu}, \bibinfo{person}{Mingxing Tan},
  \bibinfo{person}{Steve Dai}, \bibinfo{person}{Ritchie Zhao}, {and}
  \bibinfo{person}{Zhiru Zhang}.} \bibinfo{year}{2017}\natexlab{}.
\newblock \showarticletitle{Architecture and Synthesis for Area-Efficient
  Pipelining of Irregular Loop Nests}.
\newblock \bibinfo{journal}{\emph{IEEE Transactions on Computer-Aided Design of
  Integrated Circuits and Systems}} \bibinfo{volume}{36}, \bibinfo{number}{11}
  (\bibinfo{year}{2017}), \bibinfo{pages}{1817--1830}.
\newblock
\urldef\tempurl%
\url{https://doi.org/10.1109/TCAD.2017.2664067}
\showDOI{\tempurl}


\bibitem[Liu et~al\mbox{.}(2016)]%
        {splitting-2016}
\bibfield{author}{\bibinfo{person}{Junyi Liu}, \bibinfo{person}{John
  Wickerson}, {and} \bibinfo{person}{George~A. Constantinides}.}
  \bibinfo{year}{2016}\natexlab{}.
\newblock \showarticletitle{Loop Splitting for Efficient Pipelining in
  High-Level Synthesis}. In \bibinfo{booktitle}{\emph{2016 IEEE 24th Annual
  International Symposium on Field-Programmable Custom Computing Machines
  (FCCM)}}. \bibinfo{pages}{72--79}.
\newblock
\urldef\tempurl%
\url{https://doi.org/10.1109/FCCM.2016.27}
\showDOI{\tempurl}


\bibitem[Mantovani et~al\mbox{.}(2020)]%
        {hl5-2020}
\bibfield{author}{\bibinfo{person}{Paolo Mantovani}, \bibinfo{person}{Robert
  Margelli}, \bibinfo{person}{Davide Giri}, {and} \bibinfo{person}{Luca~P.
  Carloni}.} \bibinfo{year}{2020}\natexlab{}.
\newblock \showarticletitle{HL5: A 32-bit RISC-V Processor Designed with
  High-Level Synthesis}. In \bibinfo{booktitle}{\emph{2020 IEEE Custom
  Integrated Circuits Conference (CICC)}}. \bibinfo{pages}{1--8}.
\newblock
\urldef\tempurl%
\url{https://doi.org/10.1109/CICC48029.2020.9075913}
\showDOI{\tempurl}


\bibitem[Matai et~al\mbox{.}(2016)]%
        {resolve-2016}
\bibfield{author}{\bibinfo{person}{Janarbek Matai}, \bibinfo{person}{Dustin
  Richmond}, \bibinfo{person}{Dajung Lee}, \bibinfo{person}{Zac Blair},
  \bibinfo{person}{Qiongzhi Wu}, \bibinfo{person}{Amin Abazari}, {and}
  \bibinfo{person}{Ryan Kastner}.} \bibinfo{year}{2016}\natexlab{}.
\newblock \showarticletitle{Resolve: Generation of High-Performance Sorting
  Architectures from High-Level Synthesis}. In
  \bibinfo{booktitle}{\emph{Proceedings of the 2016 ACM/SIGDA International
  Symposium on Field-Programmable Gate Arrays}} (Monterey, California, USA)
  \emph{(\bibinfo{series}{FPGA '16})}. \bibinfo{publisher}{Association for
  Computing Machinery}, \bibinfo{address}{New York, NY, USA},
  \bibinfo{pages}{195–204}.
\newblock
\showISBNx{9781450338561}
\urldef\tempurl%
\url{https://doi.org/10.1145/2847263.2847268}
\showDOI{\tempurl}


\bibitem[McKeeman(1965)]%
        {peephole-1965}
\bibfield{author}{\bibinfo{person}{W.~M. McKeeman}.}
  \bibinfo{year}{1965}\natexlab{}.
\newblock \showarticletitle{Peephole Optimization}.
\newblock \bibinfo{journal}{\emph{Commun. ACM}} \bibinfo{volume}{8},
  \bibinfo{number}{7} (\bibinfo{date}{jul} \bibinfo{year}{1965}),
  \bibinfo{pages}{443–444}.
\newblock
\showISSN{0001-0782}
\urldef\tempurl%
\url{https://doi.org/10.1145/364995.365000}
\showDOI{\tempurl}


\bibitem[Mellor-Crummey and Scott(1991)]%
        {ticket-1991}
\bibfield{author}{\bibinfo{person}{John~M. Mellor-Crummey} {and}
  \bibinfo{person}{Michael~L. Scott}.} \bibinfo{year}{1991}\natexlab{}.
\newblock \showarticletitle{Algorithms for Scalable Synchronization on
  Shared-Memory Multiprocessors}.
\newblock \bibinfo{journal}{\emph{ACM Trans. Comput. Syst.}}
  \bibinfo{volume}{9}, \bibinfo{number}{1} (\bibinfo{date}{feb}
  \bibinfo{year}{1991}), \bibinfo{pages}{21–65}.
\newblock
\showISSN{0734-2071}
\urldef\tempurl%
\url{https://doi.org/10.1145/103727.103729}
\showDOI{\tempurl}


\bibitem[Morvan et~al\mbox{.}(2011)]%
        {morvan-2011}
\bibfield{author}{\bibinfo{person}{Antoine Morvan}, \bibinfo{person}{Steven
  Derrien}, {and} \bibinfo{person}{Patrice Quinton}.}
  \bibinfo{year}{2011}\natexlab{}.
\newblock \showarticletitle{Efficient nested loop pipelining in high level
  synthesis using polyhedral bubble insertion}. In
  \bibinfo{booktitle}{\emph{2011 International Conference on Field-Programmable
  Technology}}. \bibinfo{pages}{1--10}.
\newblock
\urldef\tempurl%
\url{https://doi.org/10.1109/FPT.2011.6132715}
\showDOI{\tempurl}


\bibitem[Nagarajan et~al\mbox{.}(2020)]%
        {hill-2020}
\bibfield{author}{\bibinfo{person}{Vijay Nagarajan}, \bibinfo{person}{Daniel~J.
  Sorin}, \bibinfo{person}{Mark~D. Hill}, \bibinfo{person}{David~A. Wood},
  {and} \bibinfo{person}{Natalie~Enright Jerger}.}
  \bibinfo{year}{2020}\natexlab{}.
\newblock \bibinfo{booktitle}{\emph{A Primer on Memory Consistency and Cache
  Coherence} (\bibinfo{edition}{2nd} ed.)}.
\newblock \bibinfo{publisher}{Morgan \& Claypool Publishers}.
\newblock
\showISBNx{1681737094}


\bibitem[Nickolls et~al\mbox{.}(2008)]%
        {cuda-2008}
\bibfield{author}{\bibinfo{person}{John Nickolls}, \bibinfo{person}{Ian Buck},
  \bibinfo{person}{Michael Garland}, {and} \bibinfo{person}{Kevin Skadron}.}
  \bibinfo{year}{2008}\natexlab{}.
\newblock \showarticletitle{Scalable Parallel Programming with CUDA: Is CUDA
  the Parallel Programming Model That Application Developers Have Been Waiting
  For?}
\newblock \bibinfo{journal}{\emph{Queue}} \bibinfo{volume}{6},
  \bibinfo{number}{2} (\bibinfo{date}{mar} \bibinfo{year}{2008}),
  \bibinfo{pages}{40–53}.
\newblock
\showISSN{1542-7730}
\urldef\tempurl%
\url{https://doi.org/10.1145/1365490.1365500}
\showDOI{\tempurl}


\bibitem[Nigam et~al\mbox{.}(2020)]%
        {dahlia-2020}
\bibfield{author}{\bibinfo{person}{Rachit Nigam}, \bibinfo{person}{Sachille
  Atapattu}, \bibinfo{person}{Samuel Thomas}, \bibinfo{person}{Zhijing Li},
  \bibinfo{person}{Theodore Bauer}, \bibinfo{person}{Yuwei Ye},
  \bibinfo{person}{Apurva Koti}, \bibinfo{person}{Adrian Sampson}, {and}
  \bibinfo{person}{Zhiru Zhang}.} \bibinfo{year}{2020}\natexlab{}.
\newblock \showarticletitle{Predictable accelerator design with time-sensitive
  affine types}. In \bibinfo{booktitle}{\emph{Proceedings of the 41st ACM
  SIGPLAN Conference on Programming Language Design and Implementation}}
  (London, UK) \emph{(\bibinfo{series}{PLDI 2020})}.
  \bibinfo{publisher}{Association for Computing Machinery},
  \bibinfo{address}{New York, NY, USA}, \bibinfo{pages}{393–407}.
\newblock
\showISBNx{9781450376136}
\urldef\tempurl%
\url{https://doi.org/10.1145/3385412.3385974}
\showDOI{\tempurl}


\bibitem[Nigam et~al\mbox{.}(2023)]%
        {filament-2023}
\bibfield{author}{\bibinfo{person}{Rachit Nigam},
  \bibinfo{person}{Pedro~Henrique Azevedo~de Amorim}, {and}
  \bibinfo{person}{Adrian Sampson}.} \bibinfo{year}{2023}\natexlab{}.
\newblock \showarticletitle{Modular Hardware Design with Timeline Types}.
\newblock \bibinfo{journal}{\emph{Proc. ACM Program. Lang.}}
  \bibinfo{volume}{7}, \bibinfo{number}{PLDI}, Article \bibinfo{articleno}{120}
  (\bibinfo{date}{jun} \bibinfo{year}{2023}), \bibinfo{numpages}{25}~pages.
\newblock
\urldef\tempurl%
\url{https://doi.org/10.1145/3591234}
\showDOI{\tempurl}


\bibitem[Nigam et~al\mbox{.}(2021)]%
        {calyx-2021}
\bibfield{author}{\bibinfo{person}{Rachit Nigam}, \bibinfo{person}{Samuel
  Thomas}, \bibinfo{person}{Zhijing Li}, {and} \bibinfo{person}{Adrian
  Sampson}.} \bibinfo{year}{2021}\natexlab{}.
\newblock \showarticletitle{A compiler infrastructure for accelerator
  generators}. In \bibinfo{booktitle}{\emph{Proceedings of the 26th ACM
  International Conference on Architectural Support for Programming Languages
  and Operating Systems}} (Virtual, USA) \emph{(\bibinfo{series}{ASPLOS '21})}.
  \bibinfo{publisher}{Association for Computing Machinery},
  \bibinfo{address}{New York, NY, USA}, \bibinfo{pages}{804–817}.
\newblock
\showISBNx{9781450383172}
\urldef\tempurl%
\url{https://doi.org/10.1145/3445814.3446712}
\showDOI{\tempurl}


\bibitem[Nikhil(2004)]%
        {bluespec-2004}
\bibfield{author}{\bibinfo{person}{R. Nikhil}.}
  \bibinfo{year}{2004}\natexlab{}.
\newblock \showarticletitle{Bluespec System Verilog: efficient, correct RTL
  from high level specifications}. In \bibinfo{booktitle}{\emph{Proceedings.
  Second ACM and IEEE International Conference on Formal Methods and Models for
  Co-Design, 2004. MEMOCODE '04.}} \bibinfo{pages}{69--70}.
\newblock
\urldef\tempurl%
\url{https://doi.org/10.1109/MEMCOD.2004.1459818}
\showDOI{\tempurl}


\bibitem[Ramanathan et~al\mbox{.}(2018)]%
        {concurrency-aware-2018}
\bibfield{author}{\bibinfo{person}{Nadesh Ramanathan},
  \bibinfo{person}{George~A. Constantinides}, {and} \bibinfo{person}{John
  Wickerson}.} \bibinfo{year}{2018}\natexlab{}.
\newblock \showarticletitle{Concurrency-Aware Thread Scheduling for High-Level
  Synthesis}. In \bibinfo{booktitle}{\emph{2018 IEEE 26th Annual International
  Symposium on Field-Programmable Custom Computing Machines (FCCM)}}.
  \bibinfo{pages}{101--108}.
\newblock
\urldef\tempurl%
\url{https://doi.org/10.1109/FCCM.2018.00025}
\showDOI{\tempurl}


\bibitem[Ramanathan et~al\mbox{.}(2017)]%
        {weak-consistent-hls-2017}
\bibfield{author}{\bibinfo{person}{Nadesh Ramanathan},
  \bibinfo{person}{Shane~T. Fleming}, \bibinfo{person}{John Wickerson}, {and}
  \bibinfo{person}{George~A. Constantinides}.} \bibinfo{year}{2017}\natexlab{}.
\newblock \showarticletitle{Hardware Synthesis of Weakly Consistent C
  Concurrency}. In \bibinfo{booktitle}{\emph{Proceedings of the 2017 ACM/SIGDA
  International Symposium on Field-Programmable Gate Arrays}} (Monterey,
  California, USA) \emph{(\bibinfo{series}{FPGA '17})}.
  \bibinfo{publisher}{Association for Computing Machinery},
  \bibinfo{address}{New York, NY, USA}, \bibinfo{pages}{169–178}.
\newblock
\showISBNx{9781450343541}
\urldef\tempurl%
\url{https://doi.org/10.1145/3020078.3021733}
\showDOI{\tempurl}


\bibitem[Rau and Glaeser(1981)]%
        {Rau-1981}
\bibfield{author}{\bibinfo{person}{B.~R. Rau} {and} \bibinfo{person}{C.~D.
  Glaeser}.} \bibinfo{year}{1981}\natexlab{}.
\newblock \showarticletitle{Some Scheduling Techniques and an Easily
  Schedulable Horizontal Architecture for High Performance Scientific
  Computing}.
\newblock \bibinfo{journal}{\emph{SIGMICRO Newsl.}} \bibinfo{volume}{12},
  \bibinfo{number}{4} (\bibinfo{date}{dec} \bibinfo{year}{1981}),
  \bibinfo{pages}{183–198}.
\newblock
\showISSN{1050-916X}
\urldef\tempurl%
\url{https://doi.org/10.1145/1014192.802449}
\showDOI{\tempurl}


\bibitem[Riazi et~al\mbox{.}(2020)]%
        {heax-2020}
\bibfield{author}{\bibinfo{person}{M.~Sadegh Riazi}, \bibinfo{person}{Kim
  Laine}, \bibinfo{person}{Blake Pelton}, {and} \bibinfo{person}{Wei Dai}.}
  \bibinfo{year}{2020}\natexlab{}.
\newblock \showarticletitle{HEAX: An Architecture for Computing on Encrypted
  Data}. In \bibinfo{booktitle}{\emph{Proceedings of the Twenty-Fifth
  International Conference on Architectural Support for Programming Languages
  and Operating Systems}} (Lausanne, Switzerland)
  \emph{(\bibinfo{series}{ASPLOS '20})}. \bibinfo{publisher}{Association for
  Computing Machinery}, \bibinfo{address}{New York, NY, USA},
  \bibinfo{pages}{1295–1309}.
\newblock
\showISBNx{9781450371025}
\urldef\tempurl%
\url{https://doi.org/10.1145/3373376.3378523}
\showDOI{\tempurl}


\bibitem[{Siemens EDA}(2023)]%
        {catapult-c}
\bibfield{author}{\bibinfo{person}{{Siemens EDA}}.}
  \bibinfo{year}{2023}\natexlab{}.
\newblock \bibinfo{title}{Catapult High-Level Synthesis and Verification}.
\newblock
  \bibinfo{howpublished}{\url{https://eda.sw.siemens.com/en-US/ic/catapult-high-level-synthesis/}}.
\newblock
\newblock
\shownote{(accessed: 11.1.2023)}.


\bibitem[Steele and Jay(1976)]%
        {lambda-imperative-1976}
\bibfield{author}{\bibinfo{person}{Guy Lewis~Sussman Steele, Jr.} {and}
  \bibinfo{person}{Gerald Jay}.} \bibinfo{year}{1976}\natexlab{}.
\newblock \showarticletitle{Lambda: The Ultimate Imperative}.
\newblock  (\bibinfo{year}{1976}), \bibinfo{pages}{41}.
\newblock
\urldef\tempurl%
\url{http://hdl.handle.net/1721.1/5790}
\showURL{%
\tempurl}


\bibitem[Stroustrup(1994)]%
        {design-cpp-1994}
\bibfield{author}{\bibinfo{person}{Bjarne Stroustrup}.}
  \bibinfo{year}{1994}\natexlab{}.
\newblock \bibinfo{booktitle}{\emph{The Design and Evolution of C++}}.
\newblock \bibinfo{publisher}{Addison-Wesley Professional}.
\newblock
\showISBNx{0201543303}


\bibitem[Wolfe(1986)]%
        {skewing-1986}
\bibfield{author}{\bibinfo{person}{Michael Wolfe}.}
  \bibinfo{year}{1986}\natexlab{}.
\newblock \showarticletitle{Loop Skewing: The Wavefront Method Revisited}.
\newblock \bibinfo{journal}{\emph{Int. J. Parallel Program.}}
  \bibinfo{volume}{15}, \bibinfo{number}{4} (\bibinfo{date}{oct}
  \bibinfo{year}{1986}), \bibinfo{pages}{279–293}.
\newblock
\showISSN{0885-7458}
\urldef\tempurl%
\url{https://doi.org/10.1007/BF01407876}
\showDOI{\tempurl}


\bibitem[{Xilinx Inc.}(2023a)]%
        {xilinx-abstract-parallel}
\bibfield{author}{\bibinfo{person}{{Xilinx Inc.}}}
  \bibinfo{year}{2023}\natexlab{a}.
\newblock \bibinfo{title}{Abstract Parallel Programming Model for HLS}.
\newblock
  \bibinfo{howpublished}{\url{https://docs.xilinx.com/r/en-US/ug1399-vitis-hls/Abstract-Parallel-Programming-Model-for-HLS}}.
\newblock
\newblock
\shownote{(accessed: 11.1.2023)}.


\bibitem[{Xilinx Inc.}(2023b)]%
        {xilinx-hls}
\bibfield{author}{\bibinfo{person}{{Xilinx Inc.}}}
  \bibinfo{year}{2023}\natexlab{b}.
\newblock \bibinfo{title}{Vitis HLS}.
\newblock
  \bibinfo{howpublished}{\url{https://www.xilinx.com/products/design-tools/vitis/vitis-hls.html}}.
\newblock
\newblock
\shownote{(accessed: 11.1.2023)}.


\bibitem[Zagieboylo et~al\mbox{.}(2022)]%
        {pdl-pldi2022}
\bibfield{author}{\bibinfo{person}{Drew Zagieboylo}, \bibinfo{person}{Charles
  Sherk}, \bibinfo{person}{Gookwon~Edward Suh}, {and}
  \bibinfo{person}{Andrew~C. Myers}.} \bibinfo{year}{2022}\natexlab{}.
\newblock \showarticletitle{PDL: A High-Level Hardware Design Language for
  Pipelined Processors}. In \bibinfo{booktitle}{\emph{Proceedings of the 43rd
  ACM SIGPLAN International Conference on Programming Language Design and
  Implementation}} (San Diego, CA, USA) \emph{(\bibinfo{series}{PLDI 2022})}.
  \bibinfo{publisher}{Association for Computing Machinery},
  \bibinfo{address}{New York, NY, USA}, \bibinfo{pages}{719–732}.
\newblock
\showISBNx{9781450392655}
\urldef\tempurl%
\url{https://doi.org/10.1145/3519939.3523455}
\showDOI{\tempurl}


\bibitem[Zhang and Liu(2013)]%
        {sdc-13}
\bibfield{author}{\bibinfo{person}{Zhiru Zhang} {and} \bibinfo{person}{Bin
  Liu}.} \bibinfo{year}{2013}\natexlab{}.
\newblock \showarticletitle{SDC-based modulo scheduling for pipeline
  synthesis}. In \bibinfo{booktitle}{\emph{2013 IEEE/ACM International
  Conference on Computer-Aided Design (ICCAD)}}. \bibinfo{pages}{211--218}.
\newblock
\urldef\tempurl%
\url{https://doi.org/10.1109/ICCAD.2013.6691121}
\showDOI{\tempurl}


\end{thebibliography}

\end{document}